%% file: paper.tex
\documentclass[11pt,a4paper,preprint]{article}

  \usepackage[colorinlistoftodos,backgroundcolor=white,prependcaption,textsize=tiny]{todonotes}
	
	\newcommand{\changeAP}[1]{\todo[linecolor=black!30,bordercolor=black!30]{\textcolor{black}{#1}}}
	
	\usepackage{changes}
		\definechangesauthor[color=orange]{AP}
		\definechangesauthor[color=green!55!black]{GJ}
		\setremarkmarkup{\changeAP{#2}}
		\setauthormarkuptext{}
	
\usepackage{jheppub}
\usepackage{xcolor}
\usepackage{latexsym}
\usepackage{amsmath}
\usepackage{hyperref}
\usepackage{graphicx}
\usepackage{slashed}
\usepackage{multirow}

 \def\D{\displaystyle}
 
 \def\l{\left}
 \def\r{\right}
 \def\nf{n_{\!f}}
 \def\Im{\mathrm{Im}\, }
 
 \def\d{d}
 \def\pd{\partial}
 \def\eg{e.\,g.}
 \def\ie{i.\,e.}
 \def\LO{{_{\mathrm{LO}}}}
 \def\NLL{{_{\mathrm{NLL}}}}
   \font\thin=cmbx8 
 \def\col{\thin \raise1.1pt\hbox to 0.3pt{\hss :}}
 \def\be{\begin{equation}}
 \def\ee{\end{equation}}
 \def\bea{\begin{eqnarray}}
 \def\eea{\end{eqnarray}}
 \def\bean{\begin{eqnarray*}}
 \def\eean{\end{eqnarray*}}
 \def\gsim{\mathrel{\rlap{\lower0.2em\hbox{$\sim$}}\raise0.2em\hbox{$>$}}}
 \def\lsim{\mathrel{\rlap{\lower0.2em\hbox{$\sim$}}\raise0.2em\hbox{$<$}}}
 \def\lg{\mathrel{\rlap{\lower0.25em\hbox{$>$}}\raise0.25em\hbox{$<$}}}
 \def\gl{\mathrel{\rlap{\lower0.25em\hbox{$<$}}\raise0.25em\hbox{$>$}}}

 \def\bm#1{\mbox{\boldmath$#1$}}

 \def\u#1{\underline{#1}}
 \def\wv{\widetilde}
 \def\FigScale{0.9}

 \def\bm#1{\mbox{\boldmath$#1$}}
 \newcommand{\eq}[1]{(\ref{#1})} 
  \newcommand{\bib}{
    \bibliography{\string~/lit/articles/hep,\string~/lit/books/books}
  }

\begin{document}

\title{Re-running the QCD shear viscosity}

\author[1]{Greg Jackson,%
\note{Now at {\em AEC, Institute for Theoretical Physics, University of Bern, Sidlerstrasse 5, CH-3012 Bern, Switzerland}.  }}
\emailAdd{jackson@itp.unibe.ch}
\author{Andr\'{e} Peshier}
\emailAdd{andre.peshier@uct.ac.za}
\affiliation{
  Department of Physics, 
  University of Cape Town, 
  Rondebosch 7701, 
  South Africa
}

\abstract{
  The remarkably small shear viscosity to entropy density ratio $\eta/s \lsim 0.5$ 
  of the quark-gluon plasma (QGP) is a key insight from heavy ion experiments. 
  Nonetheless, the basic understanding of this `observable' still seems to be 
  rudimentary, with existing perturbative QCD estimates suggesting $\eta/s \gsim 1$ 
  -- a view that we scrutinize here. 

  In order to extrapolate the available perturbative approach to phenomenologically 
  relevant temperatures, we consider carefully controllable higher-order corrections:
  We adapt the leading-order effective kinetic scheme 
  (instead of the catchy next-to-leading log formula 
  $\eta_{_{{\rm NLL}}} \propto [\alpha^2 \log(c/\alpha)]^{-1}$), 
  by using cross sections with a running coupling.
  This effect should be pertinent (for a QGP) and we argue for a choice
  of scale-dependence that makes it consistent with thermal screening.

  We conclude that $\eta/s \lsim 0.5$ does not indicate genuine non-perturbative 
  effects, it may rather be understood within resummation-improved perturbative QCD, 
  with some uncertainties from deploying kinetic theory near to its limit.
}

\maketitle

\input{section-1}
\clearpage
\input{section-2}

\clearpage
\input{section-3}

\clearpage
\input{section-4}
\clearpage
\input{section-5}

\section*{Acknowledgments}

G.J. was supported by the National Institute for Theoretical Physics (NITheP).

\clearpage
\appendix

\input{appendix-1}

\input{appendix-2}

\input{appendix-3}

\bibliographystyle{JHEP}
\bib

\end{document}

%% file: section-1.tex
\section{Introduction}
\label{sec:intro}


Quantifying the shear viscosity $\eta$ of the quark-gluon plasma is a key issue in
heavy-ion physics, and has been 
ever since RHIC and LHC experiments revealed it to be an (almost) perfect fluid
\cite{Teaney2003a}.
Successful descriptions of the data favour a small viscosity to entropy ratio, 
$\eta/s \lsim 0.5$: the hallmark for a `strongly coupled' system.
This feature of many-body QCD, 
along with the associated puzzle of rapid thermalisation,
is not yet properly understood and remains a challenge to explain theoretically 
\cite{Danielewicz1985}.


A strong-coupling calculation for certain supersymmetric Yang-Mills theories gives
$\eta/s=1/(4\pi)\approx 0.1$ \cite{Kovtun2005}, which appears to be in the right ballpark.
This value was conjectured as a universal lower bound for conformal gauge theories, 
but digresses from real-world QCD \cite{Huot2007}.
Moreover, it does not reveal anything about the temperature dependence of $\eta$ in a QGP.
Certain {\em models} for $\eta(T)$ can be eliminated by comparing hydrodynamic output
to transverse particle spectra and elliptic flow \cite{Niemi2015qia,Denicol2016}.
Ongoing lattice computations have helped extract $\eta(T)$ for a gluonic system near $T_c$, 
by reverse-engineering the Green-Kubo formulae.
However, the {\em assumed} form of the spectral function does not respect asymptotic freedom 
\cite{Meyer2007f, Meyer2009e, Mages2015, Astrakhantsev2017}.

First principle perturbation theory is  apparently at odds with the small viscosity,
commonly reckoned to imply that {\em all} `weak-coupling' approaches are inadequate.
Actually, this point of view is misleading as it follows from an incomplete approximation.
We show how an asymptotic weakening of the strong interaction, 
when treated consistently at the 1-loop level,
leads to a natural explanation for the low values of $\eta/s$.

\begin{figure}
   \centerline{\includegraphics[scale=1.]{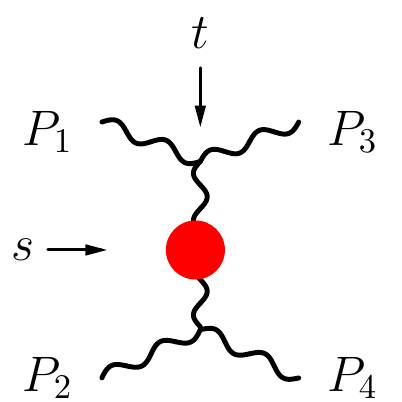}}
  \caption{
    Gluon-gluon scattering in the $t$-channel at LO produces a large forward peak in $\d \sigma / \d t$ 
    due to the long range of `glancing' interactions (a $u$-channel process contributes similarly by crossing). 
    With $P=(E,\bm p)$ denoting the four-momenta, we label the colliding partners as 
    $\{ P_1, P_2 \} \rightarrow \{ P_3, P_4 \}$.
    The exchanged gluon (carrying 4-momentum $Q=P_1-P_3$ and thus $t=Q^2$) is dressed by interactions with the medium, 
    as a minimal requirement for the LL result.
  }
  \label{1234}
\end{figure}

Kinetic theory ought to give quantitative estimates for transport properties provided 
the mean free path length $\lambda$ (specified by the associated cross section) exceeds 
the inter-particle distance \cite{Reif1964}.
The viscosity is given by ($\iota\approx 1/3$ is a numerical factor)
\be
  \eta \simeq \iota \cdot n \bar p \lambda
  \label{eta estimate}
\ee
from the density of particles $n$ which can transfer a typical momentum $\bar p$ over 
a distance $\lambda \gsim n^{-1/3}$.
For gauge theories it is important to distinguish between the total cross
section and $\sigma_{\rm tr} = \int \d\Omega\, (1-\cos\theta)\, \d\sigma/\d\Omega$.
Relevant for the viscosity is the latter, the so-called transport cross section, which
takes into account that a single one of the prevailing small-angle scatterings is not
sufficient to transfer the typical momentum $\bar p \sim T$ (for relativistic plasmas).
Although the `transport weight' $(1-\cos\theta)$, which is proportional to the
invariant momentum exchange $t$, does reduce the sensitivity of the viscosity to the 
infrared sector, it is still {\em mandatory} to also take 
into account quantum corrections in order to obtain non-trivial results.
E.\,g., for $t$-channel boson exchange (see Fig.~\ref{1234}) 
with $d\sigma/dt \sim \alpha^2/t^2$ at
tree-level, the transport cross section is logarithmically divergent,
arising from the integral $\int_{-s}^{t_{\rm max}} \d t/(-t) $.
Due to the kinematic boundary $t_{\rm max} = 0$, the
viscosity would be zero (in fact for any value of the bare coupling).
However, the exchanged boson acquires a self-energy of the order $\mu^2 \sim \alpha T^2$
due to thermal fluctuations which cannot be omitted in particular for $t \to 0\,$.
For small coupling $\alpha$, $\mu^2$ is well separated from the typical invariant
collision energy\footnote{The context will make clear the distinction between
Mandelstam $s$ and the entropy.} $s \sim T^2$. 
Thus $\sigma_{\rm tr}$ can be `mimicked' by an unscreened differential cross section,
integrated over a restricted $t$-range,
\be
  \sigma_{\rm tr}
  \sim
  \frac1s \int_{-s}^{-\mu^2} \d t\, |t| 
  \, \frac{\alpha^2}{t^2}
  \sim
  \frac{\alpha^2}{T^2} \log \l( \alpha^{-1} \r) \, .
  \label{eq1}
\ee
The emerging {\em leading logarithm} (LL) indicates the screening of soft 
scatterings due to loop corrections.
While details of the regulator $\mu^2$ do not affect the LL prefactor, 
they determine the argument of the logarithm, giving subleading terms 
${\cal O}(\alpha^2)$ in Eq.~\eq{eq1}.
We emphasize that the argument $\alpha^{-1}$ of the characteristic 
Coulomb logarithm is a ratio of `hard' and `soft' scales.

The corresponding path length is $\lambda \sim (n\sigma_{\rm tr})^{-1} \sim 1/(T
\alpha^2 \log(\alpha^{-1}))$, where we used $n \sim T^3$ for the particle density.
Since $n$ is also proportional to the entropy density, \eq{eta estimate} gives the
parametric $\alpha$-dependence
\be
  \eta/s
  \sim
  \bar p \lambda
 \sim
 1/\l( \alpha^2 \log(\alpha^{-1}) \r)
 \label{eta/s parametrically}
\ee
for hot gauge theories.


We briefly summarise the history of $\eta$ calculations: first attempts were based on
the relaxation time approximation of the Boltzmann equation \cite{Hosoya1985a}.
Those estimates were improved by Baym and collaborators who linearised the QCD
Boltzmann equation with the collision term screened with {\em hard thermal loop} (HTL)
insertions \cite{Baym1990, Heiselberg1994b}. 
This fixed the overall prefactor in \eq{eta/s parametrically}, to give the
 LL result.
As it turned out, inelastic processes (na\"{i}vely of higher order in $\alpha$) need
be included to go beyond the LL approximation.
This was recognized by Arnold, Moore and Yaffe \cite{Arnold2000d}, 
who subsequently calculated the coefficient $c$ in the {\em next-to leading log} (NLL) and
state-of-the-art result \cite{Arnold2003f}
\be
  \eta_{_{\mathrm{NLL}}}
  =
  \frac{b T^3}{ \alpha^2 \log(c/\alpha) } \, .
  \label{eta NLL}
\ee
The constants $b$ and $c$ were computed numerically, for various number of flavors $\nf\,$,
from the leading order (LO) fixed-coupling result, $\eta_{_{\mathrm{LO}}}^{\rm fix}$,
obtained in the effective kinetic framework \cite{Arnold2003a}; in the quenched limit
$b \simeq 0.34$ and $c \simeq 0.61$.


In general, $\eta(\alpha)$ is a decreasing function as expected on physics grounds: 
velocity gradients should equilibrate more efficiently by stronger interactions
(unless the quasiparticle structure of the system changes, like in phase transitions).
This expectation is indeed met by the NLL result \eq{eta NLL} for small $\alpha$, where it is
strictly justified.
However, $\eta_{_{\mathrm{NLL}}}(\alpha)$ has a minimum at $\alpha^\star = c/\sqrt{e}$
($e$ is Euler's number).
Numerically, $\min \l[ \eta_{_{\mathrm{NLL}}} \r] = 2b e T^3 / c^2$ turns out to be
close to the free entropy $s_0 = (16 + \frac{21}2 \nf) \frac{4\pi^2}{90} T^3$, 
which overestimates the entropy of the interacting QGP
(in particular near the confinement transition).
Thus it is plain that the NLL result \eq{eta NLL} is incompatible with $\eta / s \lsim
0.5$ -- which may have led to the common view that perturbative QCD cannot explain the
experimental findings.
In order to scrutinize this position, we may see the minimum at $\alpha^\star$ as a
precursor to the singularity of $\eta_{_{\mathrm{NLL}}}(\alpha)$ at $\alpha = c$,
which marks the ultimate break-down of the NLL approximation.
However, as is obvious from the derivation of the parametric formula \eq{eta/s parametrically}, 
this singularity is {\em unphysical} because it stems from a swapping of integration bounds in \eq{eq1}, 
where screening was taken into account by modifying $t_{\rm max} = 0 \to -\mu^2$.
This suggests that also the minimum of $\eta_{_{\rm{NLL}}}$ is an artefact, which will
motivate us to withhold the expansion in powers\footnote{The
situation may not improve by higher order terms in the log-expansion, because the
latter seems not to be Borel summable \cite{Arnold2003f}.} of $\log (1/\alpha)$, 
focusing instead on the full (unexpanded) LO result.


As a monotonously decreasing function, we shall find that
$\eta_{_{\rm{LO}}}(\alpha)$ could
potentially explain $\eta/s \lsim 0.5$ -- depending on the value chosen for $\alpha$.
But specifying $\alpha$ should not be guesswork.
The running of the coupling is determined by the vacuum parts of the very same 
quantum corrections whose thermal counterpart we have emphasized to be mandatory 
for screening, and thus for any reasonable approximation of transport observables 
like $\eta$.
Of course, for renormalisation-group invariant approximations, the choice of `the'
scale of the running coupling $\alpha(Q)$ is arbitrary: rescaling $Q \to \wv Q$ is
compensated by emerging $\alpha(\wv Q)\log(\wv Q/Q)$ correction terms.
These terms should either be included explicitly, or be minimized by a `natural' 
choice of scale for the relevant processes.
For the important $t$-channel process in Fig.~\ref{1234},
which leads to \eq{eq1}, this natural scale will be $Q^2 \sim t$.
Otherwise, {\em ad hoc} choices like $Q=2\pi T$ may result in
considerable inaccuracies, in particular for a quantity like $\eta$, which depends on
the square of the coupling.


%
Our paper is organised as follows.
To begin with, in Sec.~\ref{sec:sigTR}, we elaborate on the existing approximations 
and their range of validity.
From a simple model (based on the NLL result) we can make a judgement on the extrapolation 
of $\eta_\LO$ to larger values of the fixed coupling.
%
Section~\ref{sec:renorm} is where we turn QCD particularities and 
address the question of scale setting for the running coupling.
%
Our results for $\eta(T)$ are given in Sec.~\ref{sec:results},
where we also combine it with the lattice entropy to estimate $\eta/s$ for $T\sim T_c$
and then comment on kinetic theory in Sec.~\ref{sec 3C}.

%% file: section-2.tex
\section{Remarks on the existing calculation}
\label{sec:sigTR}


Screening is the basic mechanism necessary to calculate typical
transport properties, which amounts to resuming at least the
 1-loop selfenergy in the propagators.
Whether higher loop corrections will give more reliable approximations, or not, is
a relevant question, given that perturbative series do not converge.
In quantum field theories we may at best expect asymptotic
expansions\footnote{
  For QCD, Linde's argument \cite{Linde1980a} indicates further (unresolved) intricacies.}.
While not being convergent, they can still constitute useful approximations if
truncated at an appropriate `optimal' order, which usually becomes {\em smaller} with
increasing coupling strength \cite{ItzyksonZuber}.
Thus, striving for more than the 1-loop insertions absolutely necessary for screening 
may not result in more reliable approximations for transport phenomena in heavy-ion 
physics, which are characterized by a `fairly large' coupling\footnote{
  At weak coupling, higher order contribution would improve this `minimal' approximation, 
  but the relative corrections would then be small.}.
Complementing the `1-loop screening' by collinear $1 \to 2$ splitting processes defines the framework of the
LO effective kinetic theory developed in \cite{Arnold2003f} that allowed
the values $b$ and $c$ in \eq{eta NLL} to be computed.
In order to extract the LO transport properties in this effective kinetic theory 
it is sufficient to dress only the infrared sensitive scattering amplitudes for soft 
momenta.
One may then calculate transport properties for arbitrarily small $\alpha(T)$,
which (with the above motivation) make for a prudent compromise on which to
base estimates at larger coupling.
Regardless, it is the only feasible calculation scheme available at present.


Using HTL perturbation theory has the virtue of gauge invariance but 
presumes that soft and hard scales are well separated, $\mu^2 \ll T^2$.
This aspect will dominate the uncertainties when extrapolating the approach to larger coupling: 
Our estimates in Sec.~\ref{sec 2A} indicate that the HTL approximation 
gives a factor of two uncertainty for the viscosity at larger coupling. 
We may therefore simplify the approach in other aspects. 
First, we may omit the inelastic processes [which lead to ${\cal O}(5\%)$ 
higher scattering rates, thus mildly {\em lowering} $\eta$]. 
We may also solve in Sec.~\ref{sec 2B} the linearized Boltzmann equation by a 
suitable single-function Ansatz (which already gives an accuracy of $\lsim 1\%$ 
\cite{Heiselberg1994b}), instead of a full variational treatment.
The validity of a kinetic approach {\em per se} at larger coupling 
will be discussed in Sec.~\ref{sec 3C}.

\subsection{An effective model \label{sec 2A}}

Given that the non-monotonous behavior of $\eta_\NLL(\alpha)$ is related
to a sharp cut-off in the $t$-integral \eq{eq1} for the derivation of the parametric
form \eq{eta/s parametrically} of the viscosity, we are inspired to study
\bea
  \sigma^{\rm toy}_{\rm tr} (s, \mu^2)\ 
  & =  &
  \int_{-s}^0 \d t\, \frac{|t|}{2s} \, \frac{\alpha^2}{(t-\mu^2)^2}
  \ =\ 
  \frac{\alpha^2}{2s} \Big[\, \log  \l( \frac{s}{\mu^2} + 1 \r) - \frac{s}{s+\mu^2} \,\Big] ,
  \label{m1}
\eea
using a `self-energy' $\mu^2 \sim \alpha T^2$ instead of a sharp cut-off.
Earlier we tentatively assumed that the typical invariant energy for 
the transport cross section in a thermal medium  is hard, $s \sim T^2$ 
(which seems to be a reasonable view in the given context).

\begin{figure}[!htb]
  \includegraphics[scale=\FigScale]{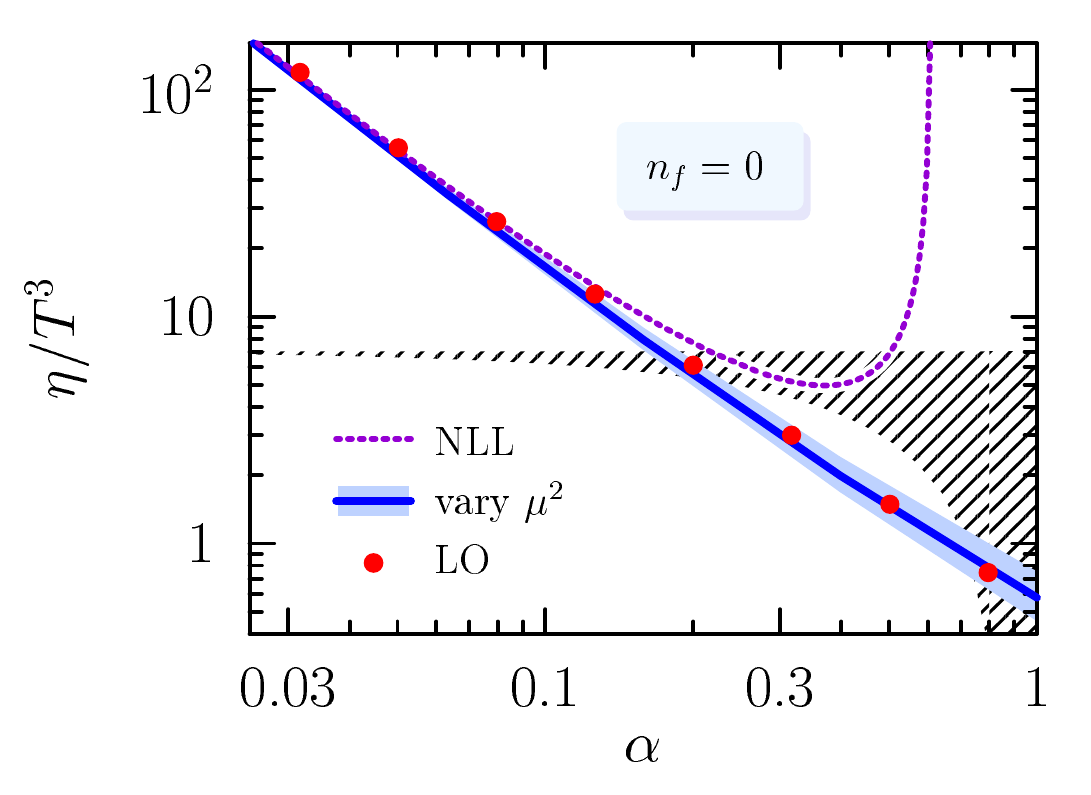}
  \caption{ 
  The QCD shear viscosity in the quenched limit as a function of fixed coupling strength, 
  in NLL approximation \eq{eta NLL} and to LO \cite{Arnold2003f}. 
  The latter is well reproduced by the model \eq{m1}, \eq{eta with P(s)}, 
  which allows to estimate the sensitivity on the screening of hard interactions, 
  see text.
   For comparison, the hatched region demarcates the entropy between the LO result 
   $s_{_{\rm LO}} = s_0 (1-\frac{15}{4\pi}\alpha)$ and the free limit.
}
  \label{fig: eta(alpha)}
\end{figure}

However, the gist of the somewhat technical framework outlined in Section~\ref{sec 2B} 
is that the viscosity is approximately a convolution of the screened transport cross 
section with a positive function which, with appropriate normalization, can be interpreted 
as a probability for momentum exchange between laminar layers,\footnote{
We will define $P(s)$ later in Eq.~\eq{Pdef}, and give some of its properties in Appendix \ref{A5}.}
\be
  \frac\eta{T^3}
  \simeq
  1 \Big/ \int ds\, sP(s) \cdot \sigma_{\rm tr}(s) \, .
  \label{eta with P(s)}
\ee
The simple scheme \eq{m1}, \eq{eta with P(s)} will allow us to discuss several aspects 
of extrapolations to larger coupling strength. 
By matching it for weak coupling to the perturbative QCD result \eq{eta NLL}, we can readily infer the normalisation $2\int ds P(s) = b^{-1}$.
In order to match also the constant $c$ in \eq{eta NLL} we adjust\footnote{
  Note that this differs from \cite{Jackson2017a}, where we simplified further and 
  took $\sigma_{\rm tr}$ at the `mean' $\bar s$, adjusted to agree with the NLL result. 
  Because $P(s)$ is monotonically decreasing, this is not fully justified and we are more 
  careful here. However, the results hardly differ, thus the simple formula in 
  \cite{Jackson2017a} might still be useful for quick estimates.
}
the regulator $\mu^2 = \kappa \cdot 4\pi\alpha T^2$: From \eq{moments of P(s)} we find 
$\kappa \simeq 0.38/c \simeq 0.62$, \ie\ $\mu^2$ is somewhat smaller than the Debye 
mass-squared, $m_D^2 = 4\pi\alpha T$.
This procedure not only takes into account the remaining elastic contributions in the 
binary QCD cross section ($t$ and $u$ channels contribute equally by crossing symmetry), 
but also inelastic processes (which give only a small correction, as noted before).
Figure \ref{fig: eta(alpha)} shows that our effective model agrees well the LO QCD result 
\cite{Arnold2003f} (reviewed in the next section) even for large coupling well beyond 
the breakdown of the NLL result \eq{eta NLL}. 

Having specified the model, we can now estimate the uncertainties that must be faced in 
QCD when using HTL-dressed propagators\footnote{
  Which are justified for $|t| \lsim T^2$ \cite{Peshier1998}, although often 
  derived under the stricter assumption that external energy and 3-momentum are 
  smaller than $T$.
},
by studying the sensitivity of the viscosity \eq{eta with P(s)} on the `self-energy' 
at {\em harder} momentum transfers.
Varying $\mu^2$ by factors $\nu = 2^{\pm 1/2}$ for $|t|>T^2$ has only a mild effect: 
Even for $\alpha \sim 1$ the viscosity changes by less than a factor of $\nu^{-1}$, 
cf.~Fig.~\ref{fig: eta(alpha)}.

Alternatively, we could also segregate momentum transfers at $t^\star \in [-s,0]$, 
and set $\mu^2 \to 0$ for $|t|>|t^\star|$, \ie\ use the Born cross section for 
harder interactions, which is basically\footnote{
  Usually a non-covariant separation in 3-momentum is made, 
  but we find it more convenient to separate in invariant momentum transfer.}
the Braaten-Yuan method \cite{Braaten1991d},
\be
\int_{-s}^{t^\star} \d t \frac{|t|}{2s} \frac{\alpha^2}{t^2}
\ +\ \sigma_{\rm tr}^{\rm toy}\big(-t^\star,\mu^2\big) 
= \frac{\alpha^2}{2s}
\Big[ \,
\log \Big(\, \frac{s}{\mu^2} - \frac{s}{t^\star}\, \Big)
- \frac{t^\star}{t^\star - \mu^2} \, \Big] \, .
\label{mu2 with t^star}
\ee
The resulting viscosity is $t^\star$-independent only at NLL order. 
The sub-leading terms lead to a $t^\star$-dependence which becomes more 
pronounced for increasing $\alpha$. Varying $|t^\star|$ in the range 
$[\frac12,2]T^2$, we observe a similar sensitivity as above on $\nu$, 
leading us to estimate that the HTL approximation does not bring more than a 
factor of two uncertainty for the viscosity calculation.

\bigskip

\begin{figure}[htb]
  \includegraphics[scale=\FigScale]{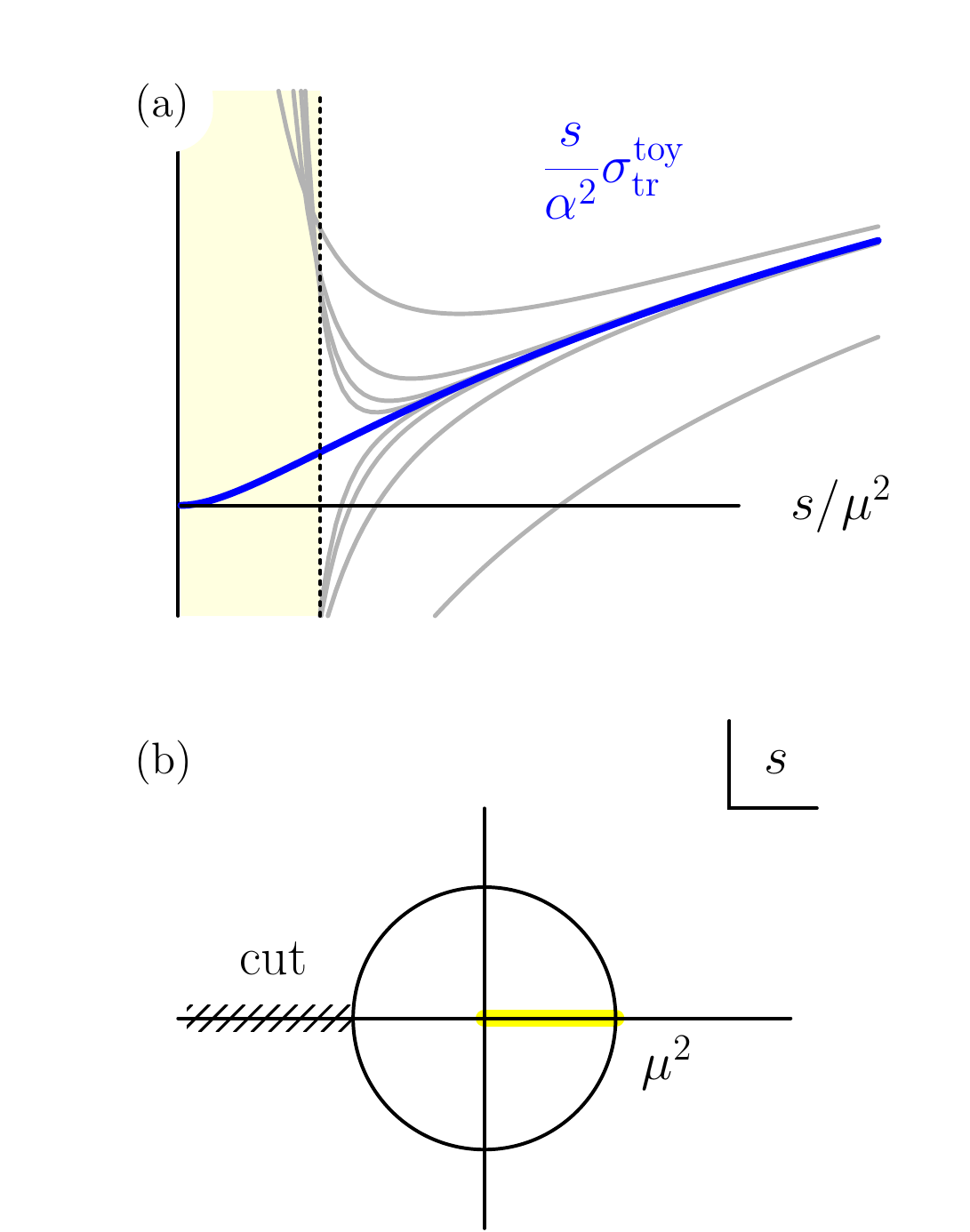}
  \caption{ 
  Panel (a) shows how the transport cross section in \eq{m1} depends on $s/\mu^2$.
  The NLL approximation, $\sigma_{\rm tr}^{\rm NLL} \sim \log (s/\mu^2)$ is accurate for $s \gg \mu^2$, 
  guaranteed for weak coupling $\alpha \to 0$.
  However an expansion in $\mu^2/s \sim \alpha$ extrapolates poorly (family of gray curves)
  for `soft' $s$, where $\sigma_{\rm tr} \approx \alpha^2 s/(2\mu^4)$.
  In the complex $s$-plane, $\sigma_{\rm tr}$ has a branch cut on the negative real axis for $s<-\mu^2$.
  Thus, shown in (b), the radius of convergence is $\mu^2$.
}
  \label{fig: wtr(s)}
\end{figure}

While the full (unexpanded) RHS of \eq{m1} is obviously a manifestly
positive, monotonously decreasing function of $\mu^2/s \sim \alpha$,
its NLL approximation, $\sigma_{\rm tr}^{_{\mathrm{NLL}}} / \alpha^2 = s^{-1} [ \log(s/\mu^2) - 1]$, 
becomes negative for $s/\mu^2 < e$, leading to the same unphysical behavior as seen in \eq{eq1} and \eq{eta NLL}. 
This problem is not cured by higher order terms since the expansion has a radius of convergence $\mu^2/s =1$, 
stemming from the pole of ${\cal M} \sim \alpha/(t-\mu^2)$ at $t=\mu^2$ (off the physical sheet) in the
integrand of \eq{m1}. 
The factor of $|t|$ cedes a branch cut on the negative real axis $s<-\mu^2$,
 rather than a pole at $s=-\mu^2$ in $\sigma = \int \d t\, \d \sigma / \d t$,
and expansions only converge in the disk of radius $\mu^2$, illustrated
in Fig.~\ref{fig: wtr(s)}.
Therefore the powers of $\mu^2/s$, formally beyond LO, 
should be considered carefully when extrapolating to larger coupling.

To substantiate this statement,
and deferring proof to Appendix \ref{A5}, we point out that $P(s)$ is non-zero for $s \to 0$.
Thus, expanding $\sigma_{\rm tr}$ in $\mu^2/s \sim \alpha T^2/s$ {\em before} convoluting
it with $P(s)$ in Eq.~\eq{eta with P(s)} yields an ill-defined series in powers of $\alpha$: 
its coefficients [the negative moments of $P(s)$] are IR-divergent, with increasing severity.
Interchanging expansion and integration is formally not permissible given the
properties of the particle's local equilibrium distribution encoded in $P(s)$.
This issue is generic; the non-monotonous behavior of
$\eta_{_{\mathrm{NLL}}}(\alpha)$ is a mathematical {\sl artifact} rather than a
breakdown of perturbative QCD {\em per se}.
It also means that these higher order by-products must be kept.

\subsection{Linearised Boltzmann equation \label{sec 2B}}


%
Having elaborated on the quality of various approximations and their respective drawbacks, 
we turn to the rigorous kinetic theory for QCD.
What was established in Refs.~\cite{Baym1990,Heiselberg1994b,Arnold2000d,Arnold2003f}
can guide us to \eq{eta with P(s)}, which we explain here for the quenched limit ($\nf =0$).

We characterise the system by $f(\bm p, \bm x, t)$,
the number of partons in the phase-space element $\d^3 \bm p \d^3 \bm x /(2\pi)^3$.
The evolution of $f$ can be described by the Boltzmann equation,
\be
{\cal D} f = {\cal C} \l[ f \r] \, ,
  \label{C[f]}
\ee
where the streamline derivative ${\cal D} = \pd_t + \bm v \bm \nabla$ for relativistic particles.
For binary scatterings (see Fig.~\ref{1234}) the collisional operator reads
\bea
{\cal C} \l[ f_1 \r] ( p_1 )
  &=&
  \frac{1}{2E_1} 
  \int \d \Gamma
  \label{C}
  \l| {\cal M} \r|^2
  \big\{\ \bar{f}_1 \bar{f_2}f_3f_4 - f_1 f_2\bar{f_3}\bar{f_4}\ \big\}  \, ,\nonumber
\eea
using the subscript on the distribution function to refer to its momentum argument, 
$f_i = f(\bm p_i)$.
Quantum effects are included by the bracketed term \eq{C}; we use the notation $\bar{f} = 1\pm f$, 
with the upper sign for bosons and the lower for fermions.
Two-body phase space is abbreviated
\bean
  \int \d \Gamma \
  &\equiv& \ 
  d_2 \int_{234} 
  \frac{ (2\pi)^4 }{8 E_2 E_3 E_4}
  \delta^{(4)} \l( P_1 + P_2 - P_3 - P_4 \r)  \, ,
\eean
using the shorthand $\int_i = \int \d^3 p_i/(2\pi)^3$; with $d_g = 16$ and $d_q = 12 \nf$
being the gluon and quark degeneracies for $\nf$ light flavours respectively. 
Note that $\int \d \Gamma$ is a Lorentz invariant measure with $P = (E, \bm p)$ being
the usual 4-momentum. 
In \eq{C}, the matrix element $\vert {\cal M} \vert^2$ is averaged
over spin and colour of incoming particles $\{1,2\}$,
but summed over final states $\{3,4\}$ (taking into account double counting).
In vacuum at Born level, ${\cal M}$ is dependent only on the Mandelstam
invariants [we use the ${\rm Diag}(1,-\bm 1)$ convention for the metric]
\bean
  s = (P_1 + P_2)^2 \, , \quad
  t = (P_1 - P_3)^2 \, , \quad
  u = (P_1 - P_4)^2 \, . 
\eean


A heat bath in equilibrium (which is static and uniform in $\bm x$)
has a collective flow and is characterised\footnote{
  We consider only zero chemical potential.} 
by a temperature $T$ and velocity $\bm u$. The J\"{u}ttner distribution function
\bea
f_{\rm eq} \l( \bm p \, ; T, \bm u \r) &=& \l( e^{\D\gamma (p - \bm p \bm u )/T } \mp 1 \r)^{-1}
\, , \label{global}
\eea
where $\gamma = \l( 1- u^2 \r)^{-1/2}$, 
solves \eq{C[f]} trivially since ${\cal C} \l[ f_{\rm eq} \r] = 0$ .
Consider now small deviations from the {\em global equilibrium} in \eq{global},
by allowing for $\bm u$ to weakly depend on $\bm x$ and consider as an {\rm Ansatz} to solving
\eq{C[f]} the function
$f(\bm x, \bm p) = f_{\rm eq} \bm{(} \bm p\, ; \bm u(\bm x) \bm{)}$,
 the so called `local' equilibrium [to be distinguished from the general
off-equilibrium solution, $f_\star$].
Small gradients in the collective flow $\bm u$ allow us to assume 
also that $|\bm u| \ll 1$, by boosting into the local rest frame.
This implicit dependence of $\bm u$ on $\bm x$ means that $f$ no longer 
satisfies the Boltzmann equation \eq{C[f]}, since\footnote{
  Suppose a steady solution, \ie\ $\pd_t \bm u \to 0$, implying $\nabla T = 0$.
  However a nonzero $\pd_t T$ is connected to any divergence in $\bm u$, 
  and is necessary for the second term in \eq{S}.
}
\bea
{\cal D} f_{\rm eq} \bm{\big(} \bm p \, ; \bm u (\bm x) \bm{\big)}
&=&
S\, \Big( \hat{p}_i \hat{p}_j - \frac13 \delta_{ij} \Big) \cdot \nabla_i u_j \, ,
\quad {\rm with}  \quad 
S(p)\ \equiv\ f \bar{f} \frac{p}{T} \, ,
\label{S}
\eea
while the collision operator in the Boltzmann equation \eq{C[f]}, still vanishes.
Thus, the actual solution to \eq{C[f]} should somewhat depart from local equilibrium,
\bean
  f_\star ( \bm x, \bm p ) &=& 
  f \big( \bm x  , \bm p \big) + \delta f (\bm p )\, ,
\eean
with $\delta f$ being proportional to the generic velocity gradient $\nabla_i u_j$, in
order for the collision operator ${\cal C} [f_\star]$ to compensate \eq{S}.
Let us parametrise $\delta f$ in the form, similar to \cite{Baym1990},
\bean
\delta f &=& f \bar{f}\ \frac{ \chi^{ij} (\bm p) }{T}
\cdot \nabla_i u_j\, ,
\eean
where the rank-2 traceless tensor $\chi^{ij}$ is 
\bea
\chi^{ij} &=& \chi (p)
\Big( \hat{p}^i \hat{p}^j - \frac13 \delta^{ij} \Big) \, ,
  \label{chiij}
\eea
in terms of a  scalar function $\chi (p)$. 
The tracelessness of $\chi^{ij}$ will imply zero bulk viscosity, isolating only the shear modes.

The deviation $\delta f$ from local equilibrium (alternatively $\chi$), 
leads to a modification of the stress tensor from the non-interacting limit,
\bean
{\cal T}_{ij}^\star
&=& d_g \frac{\pi^4}{90}T^4 \delta_{ij} +
{\cal T}_{ij} [\delta f ]
\eean
From \eq{chiij}, the first order correction (in laminar gradients) takes the conventional form
\bea
d_g \int_p \frac{p_i p_j}{p} \delta f (\bm p ) 
&=&
  - \eta \cdot 
  \Big( \,
  \nabla_i u_j + \nabla_j u_i - \frac23 \delta_{ij} \bm \nabla \bm u 
  \,   \Big) \, ,
  \label{Tij}
\eea
 where the shear viscosity is now derivable from $\chi$ :
\bea
\eta &=& \frac{1}{15} d_g \int_p \chi S \, .
\label{eta=}
\eea
But before formula \eq{eta=} may be of any use, the unknown function $\chi$ needs to be
determined -- 
under the simplifying assumption of small gradients in velocity.
The left hand side of the Boltzmann equation \eq{C[f]}, which controls the effective particle
rate, is (to first order in the gradients) ${\cal D}(f + \delta f) \simeq
{\cal D} f$, \ie\ the convective derivative is approximated by \eq{S}.
On the other hand,  Ansatz \eq{chiij} gives
\bea
{\cal C} \l[ f_1 \r]
&\simeq&
\frac{ \nabla_i u_j}{2TE_1} \int\! \d \Gamma
\vert {\cal M} \vert^2 f_1 f_2 \bar{f_3} \bar{f_4} 
\cdot \Delta^{ij} \l[ \chi_1 \r]\, ,
\label{RHS}
\eea
where (having kept terms proportional to the gradient)
\bean
\Delta^{ij} \l[ \chi_1 \r]
\ &\equiv&\ 
\big\{\ \chi_1^{ij} + \chi_2^{ij} - \chi_3^{ij} - \chi_4^{ij}\ \big\} \, .
\eean
Since $\nabla_i u_j$  was arbitrary, we may use the Boltzmann equation 
to match \eq{S} with \eq{RHS} and  hence replace $\nabla_i u_j$ by $\hat{p}_i \hat{p}_j$ to find
\bea
S(p) &=& {\cal C}_L \l[ \chi \r] (p) \, .
\label{linBE}
\eea
In this relation, the {\em linearised} collisional operator is defined by
\bean
{\cal C}_L \l[ \chi_1 \r]
&\equiv& 
\frac{3\hat{p}_1^i \hat{p}_1^j}{4E_1} \int\! \d \Gamma 
\vert {\cal M} \vert^2 f_1 f_2 \bar{f_3} \bar{f_4} 
\cdot \Delta^{ij} \l[ \chi_1 \r]\, ,
\eean
where $\chi_i = \chi(p_i)$ is an arbitrary scalar function [we reserve $\chi^\star$ for the actual solution of \eq{linBE}].
Note that $\Delta^{ij}[\chi]$ is traceless, a property inherited directly from \eq{chiij}, and therefore 
\bea
\int_p \chi {\cal C}_L \l[ \chi \r]
&=& 
\frac38 
\int_1 
\frac{f_1}{2E_1} \int\! \d \Gamma 
\vert {\cal M} \vert^2 f_2 \bar{f_3} \bar{f_4} 
\cdot \Big( \Delta^{ij} \l[ \chi_1 \r] \Big)^2 \, ,
\label{symC}
\eea
where symmetry of the integrand was used to complete the square for $\Delta^{ij}$.
Therefore ${\cal C}_L$ is a positive semidefinite operator, over the Hilbert space of $\chi$-functions, 
which vanishes only for collisionally conserved quantities.


%
Eq.~\eq{linBE} then determines $\delta f$ [via $\chi(k)$], enabling \eq{eta=} to be used
formally with the {\em exact} solution: $\chi^\star = {\cal C}_L^{-1} \l[ S \r]$.
One approach to solving this symbolic equation is to represent $\chi$ by a
linear combination over some complete set of functions.
Since ${\cal C}_L$ is linear, this would produce an (algebraic) matrix equation which
may be solved for the coefficients.
Such a strategy, using a truncated basis, 
gives a {\em lower} estimate for $\eta$ as we now explain.

\bigskip
Consider the quadratic functional, following \cite{Arnold2000d}, 
\bea
{\cal Q} \l[ \chi \r] &=&
d_g \int_p \Big( S \chi - \frac12 \chi {\cal C}_L \l[ \chi \r] \Big) \, .
\label{Q}
\eea
At $\chi^\star$, the solution to \eq{linBE} , ${\cal Q}$ takes the value $\eta$ (up to a prefactor).
The linearised Boltzmann equation is tantamount to 
\bean
\frac{\delta {\cal Q}}{\delta \chi} &=& 0  \, ,
\eean
and is solved by $\chi^\star$, where ${\cal Q}$ is stationary.
Because ${\cal C}_L$ is positive definite, see \eq{symC}, the extremum of \eq{Q}  at $\chi = \chi^\star$ is in fact a {\em maximum}. 
Its value here, on account of Eq.~\eq{eta=}, gives the viscosity:
\bea
\eta &=&  \frac{2}{15} {\rm Max} \big( {\cal Q} \big) \, ,
\label{etaVAR}
\eea
which sidesteps the task of actually inverting ${\cal C}_L$.
For example, the optimal norm of any test function may be derived from $\pd_A {\cal Q} \l[ A \chi \r]=0$, 
which in a sense `homogenises' \eq{Q}. 
We may then substitute ${\cal Q}$ with the following\footnote{
  This functional may also be obtained by a Cauchy-Schwarz inequality.} 
(improved) estimate,
\bea
{\cal Q}[\chi] &=& 
\frac{\D \Big( d_g \int_p S\chi \Big)^2}{\D 2 d_g \int_p \chi {\cal C}_L \l[ \chi \r]} \, .
\label{baym}
\eea
If \eq{baym} is used for ${\cal Q}$ instead of Eq.~\eq{Q}, 
the absolute scale of $\chi$ is arbitrary (evidently it cancels out in the fraction).
This functional, along with the known QCD matrix element for gluon-gluon scattering 
(with IR terms aptly screened), enables $\eta$ to be calculated accurately from a suitably adjusted test function $\chi$.
Appending the quark sector is a technicality which allows for `mixing' between boson and fermion distributions 
(E.g. due to fusion reactions like $gg\leftrightarrow q\bar{q}$).
To evaluate $\eta$ for a general $\nf>0$, we refer the reader to the cross section listing in Table III of Ref.~\cite{Arnold2003f}.
Numerical determination of $\int_p \chi {\cal C}_L [\chi]$ involves the phase space integration 
over all collisional momenta subject to energy and momentum conservation.
See Appendix \ref{A2} for our covariant manipulation of the resulting 5-fold integral
(which in general cannot be reduced further).

\subsection*{Connecting to Eq.~\eq{eta with P(s)}}

In order to validate our effective $t$-channel model in Sec.~\ref{sec 2A}, 
we show how Eq.~\eq{baym} leads naturally to \eq{eta with P(s)}.
The prevalent small-$t$ contribution to \eq{symC}, which we organise\footnote{
  The order of integration is modified from \eq{g's}, so that
  $s$ now imposes a restriction on the energy integrations in $P(s)$.
}
in Appendix \ref{A2}, is
\bean
\int_p \chi {\cal C}_L \l[ \chi \r] &=&
N \int_0^\infty \d s \, s P(s) 
\int_{-s}^0 \d t  \frac{|t|}{2s} \frac{\d \sigma}{\d t} 
+ \ldots \, ,
\eean
through a positive weight $P(s)$ that depends on how the system departs from equilibrium.
($N=3/8$ is just a numerical factor.)
This formula assumes that $\d \sigma / \d t$ depends only on $s$ and $t$, and only accounts 
for the dominant contribution from small-angle binary scatterings.
We note that $P$ specifies the `typical' momentum $\bar p$ in \eq{eta estimate} more rigorously.
Presuming small-$t$ dominance allows a further integration to be performed, leaving
us with $P$ in terms of an integral over the incoming energies;
\be
 P [\chi] (s) \ \ \col{=}\ 
 \int_{12} \frac{f_1\bar f_1 f_2 \bar f_2}{4E_1E_2} 
\Big(  {\cal B} + 
\frac{s}{4E_1E_2} \big[ \, 8 {\cal A} - {\cal B}\, \big]  \Big) \, ,
\label{Pdef}
\ee
where $\int_i = \int \d E_i \, E_i^2 /(2\pi^2)$ such that $4E_1E_2>s$.
And
\bea
{\cal A} &=&
 \l(\hat{\chi}_1 - \hat{\chi}_2 \r)^2 
 + \frac{s}{E_1E_2} \hat{\chi}_1 \hat \chi_2  \, , \nonumber \\
{\cal B} &=&
 \frac43 \big[\, \chi_1^\prime - \chi_2^\prime \,\big]^2 
+ 4 s \hat{\chi}^\prime_1 \hat{\chi}^\prime_2
- \frac{s^2}{E_1E_2}
\Big[\,
\hat{\chi}_1^\prime - \frac{\hat \chi_1}{E_1}
\, \Big] \Big[ \,
\hat{\chi}_2^\prime - \frac{\hat \chi_2}{E_2}
  \, \Big] \, , \nonumber
\eea
where $\hat{\chi}_i = \chi_i / E_i$ and $\hat{\chi}^\prime_i = (\pd \hat{\chi}_i/\pd E_i)$\, .

As crucial to our argument that the log expansion diverges, the value of $P(s=0)$ must be strictly positive.
This follows from ${\cal B}>0$ for $s \to 0$ in the integrand of Eq.~\eq{Pdef}, without even knowing $\chi$.
That still leaves open the possibility that $P(0)$ is finite {\em or} $P(s)$ 
has an (integrable) divergence at $s=0$.
Either is enough assurance that all the negative moments diverge -- in fact, 
even polynomial growth of $P(s)$ for small-$s$ would only tame finitely many such terms.

To validate \eq{eta with P(s)} against $\frac{2}{15}{\rm Max}[{\cal Q}]$,
we may just assign
\bea
1 &=& N\frac{d_g}{15} \l( \int_p \chi S \r)^2 \, . \label{B}
\eea
This merely reinterprets the existing NLL result, where
we could then adapt the effective cross section to resemble QCD.
In appendix \ref{A5} we show how normalisation of $P(s)$ and \eq{B} are sufficient to 
determine $\chi$ from a differential equation.
However, it turns out that $\chi(p)=p^2$ gives the stationary 
point in ${\cal Q}[\chi]$ quite accurately \cite{Baym1990}.
[We discuss why, and the arising $P(s)$ in appendix \ref{A5}.]

%% file: section-3.tex
\section{Scale(s) for screening}
\label{sec:renorm}


Evidently it may be possible to explain $\eta/s \lsim 0.5$ on the basis of a leading order
treatment (see Fig.~\ref{fig: eta(alpha)}), but to do so requires a proper understanding
of how `the' value of the coupling is specified by the temperature.
To answer this question, we need to bring up a main feature of QCD.
That will resolve the `loose end' of having to {\em impose} $Q_T\sim T$ as the relevant
scale -- something already disputed in Sec.~\ref{sec:intro}\,.
The toy cross section considered in the previous section offers some guidance for gauge theories, 
where $\mu^2$ is supplanted, \eg\ by the gluon self-energy $\Pi (Q)$ [a non-trivial function of both 
components of the four-momentum $Q=(\omega, \bm q)$\,].

The tree level cross section is [averaged over initial, summed over final and multiplied by $\frac12$ 
to remove double counting: see \eq{C}]
\be
  \frac{\d \sigma}{\d t}^{\rm tree}
  =
   \alpha^2 \frac{9\pi}{4s^2} 
   \Big[ \, - \frac{su}{t^2} - \frac{ts}{u^2} - \frac{tu}{s^2} + 3 \, \Big] \, .
   \label{glue-glue}
\ee
The first two terms in square brackets, of the type we focused on in Sec.~\ref{sec:sigTR}, 
contribute equally to $\eta$  due to $t \leftrightarrow u$ crossing.
What remains, \ie\ $-tu/s^2 + 3$, only contributes at NLL-order because they are overwhelmed 
by the first two terms for small-$t$.
A dressed gluon propagator $D=(D_0^{-1}-\Pi)^{-1}$ must be used for these infrared sensitive terms,
amounting to the replacement
\be
-su/t^2 \to \big| D_{\mu\nu}(Q) Y^{\mu\nu} \big|^2 + \tfrac14 \, ; \quad Q=P_1-P_3 \, ,
\label{IR replacement}
\ee
where $Y^{\mu\nu} = (P_1 - \frac12 Q)^\mu (P_2 + \frac12 Q)^\nu$.
The polarisation tensor $\Pi$ has a isolated finite-$T$ contribution,
(denoted here by a tilde)
\bea
\Pi_{\mu\nu} (Q) &=&
\Pi_{\mu\nu} \Big|_{T=0} + \widetilde{\Pi}_{\mu\nu} \, .
\label{Pi}
\eea
Running of the coupling $\alpha (\cdot)$ emerges from vacuum fluctuations in a process like \eq{glue-glue}, 
after the bare parameters in the Lagrangian are expressed by renormalised ones.
Generically, many types of corrections are relevant: vertex dressing, ghost contributions, etc. \cite{ItzyksonZuber}.
Only $\Pi$ is needed for coupling constant renormalisation in Coulomb gauge, 
due to its Abelian-like Ward identities \cite{Grozin2008}.
In this gauge, the (resummed) gluon propagator separates into longitudinal ($L$) and transverse ($T$) components
\bean
D_{00} \ = \ \frac{-Q^2}{\bm q^2 } \Delta_L \ ,
\qquad
D_{ij} \ = \ -\Big(\, \delta_{ij} -   \hat{q}_i \hat{q}_j \, \Big) \Delta_T  \, ,
\qquad
D_{0i} \ = \ D_{i0} \ = \ 0 \, ,
\eean
with $\Delta_{L,T} (Q) = \big(Q^2 - \Pi_{L,T}(Q) \big)^{-1}$ 
being ordinary scalar propagators\footnote{
  The scalar functions $\Pi_{L,T}$ 
are related to \eq{Pi} by
$$
\Pi_L  =  \frac{-Q^2}{\bm q^2} \Pi_{00}
\ , \qquad 
\Pi_T  =  \frac12 \Big( \hat{q}_i \hat{q}_j - \delta_{ij} \Big)  \Pi_{ij} \, .
$$
}.
$\Delta_T$ and $\Delta_L$ coincide for $T=0$, but are different at non-zero temperature.
Before returning to the issue of specifying $\alpha$, let us discuss the thermal self-energy for QCD in the quenched limit 
and see how Eq.~\eq{mu2 with t^star} translates in the full LO calculation.

\subsection{Thermal Screening\label{sec 3A}}

Coulomb gauge is customary at $T>0$ for another reason, 
namely due to the manifestly broken Lorentz invariance by the presence of a heat bath, see Ref.~\cite{Weldon1982aq}.
Self-energies in a thermal medium acquire an additional finite contribution $\widetilde{\Pi}_{\mu\nu}$, 
scaling with $T^2$ (as opposed to $Q^2=\omega^2-\bm q^2$).
But rather than being `constant' in $Q$, as $\mu^2$ was, $\widetilde{\Pi}_{\mu\nu}$ also depends separately on $\omega$ and $\bm q$\,, 
with an analytic expression valid near the light-cone; $|\omega^2-\bm q^2| \lsim T^2$ (\ie\ the HTL limit).

Parametrically, the scalars $\wv{\Pi}_{L,T}$ are proportional to $\alpha$ and may be written 
$\widetilde\Pi = m_D^2 \phi (\omega, \bm q)$ where 
$m_D^2 = (1+\frac16 \nf)4\pi \alpha T^2$ is the Debye mass.
(Equivalently, the function $\phi$ could be written in terms of the virtuality $Q^2$ and $z=\omega/|\bm q|$. 
Then $\mu^2$ is interpreted as an `average' over the argument $z$ to represent the effect of Landau damping.)
Details of $\phi_{L,T}$ are immaterial for the present discussion,
save that they are {\em finite} and therefore do not affect renormalisation.
In a fixed-$\alpha$ result, we may simply drop the vacuum contribution to $\Pi$, 
and that is what has been done previously.

It is sufficient to use HTL propagators $D^\star$ in \eq{IR replacement} for
LO accuracy because their domain of validity coincides with what is required.
HTL screening is justified for $|Q^2| \lsim T^2$ and gives $ \phi = \phi^\star(z)$ 
(details in Appendix \ref{A4}).
Subsequently, $\eta$ will be sensitive to a class of higher order corrections which arise 
because $\phi$ is not known for harder momenta.
To estimate the relevance of these subleading terms, let us again adapt the Braaten-Yuan \cite{Braaten1991d} approach 
 by omitting screening for $|Q^2| > |t^\star|$,
\ie\
$$
\phi (Q^2,z) = \Theta ( Q^2 -t^\star) \cdot \phi^\star(z) \, .
$$
The $t^\star$-dependence cancels under weak-coupling assumptions (cf. Appendix \ref{A4}).
We thus expect our conclusions about the expansion to carry through {\em mutatis mutandis} for QCD.

\begin{figure}[hbt]
  \includegraphics[scale=\FigScale]{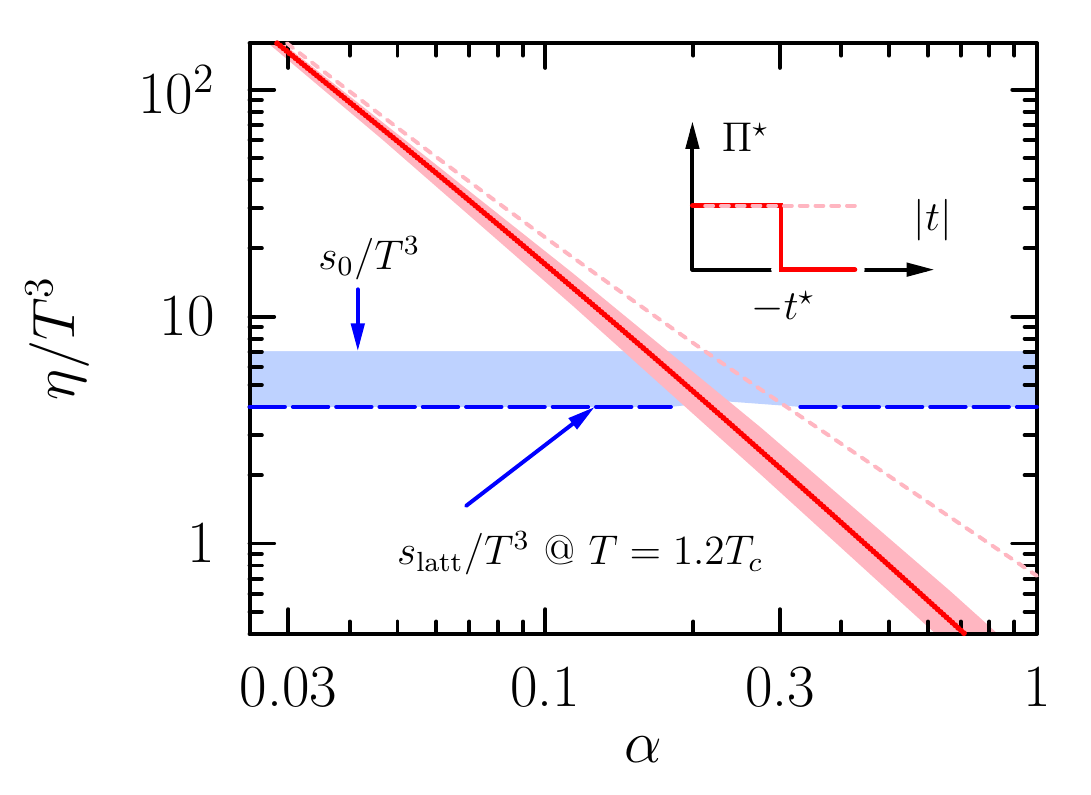}
  \caption{ 
  The viscosity for quenched QCD to LO accuracy,
  cf. Fig.~\ref{fig: eta(alpha)}.
  The curves here are all equally valid at LO, and calculated
  using HTL self-energies.
  Multiplying the self-energies by $\Theta (t-t^\star)$
  and varying $t^\star$ near the canonical value $-T^2$ 
  (as used in Ref.~\cite{Arnold2003f}, giving the `LO' points in Fig.~\ref{fig: eta(alpha)})
  produces the pink band.
  We also show the constraint for the entropy, $4T^3 \leq s \leq s_0$ for $T>1.2T_c$, 
  to confirm that $\eta_\LO$ could indeed explain how $\eta/s \lsim 0.5$.
}
  \label{fig: eta HTL}
\end{figure}

\bigskip
Therefore we use the functional ${\cal Q}$ \eq{baym}, 
and probe the residual sensitivity to the covariant cut-off $t^\star$ by varying $|t^\star| \in [\frac12, 2]T^2$.
This is illustrated in Fig.~\ref{fig: eta HTL} (the findings are similar to Sec.~\ref{sec 2A});
for $\alpha = 0.4$ there is a factor $2$ uncertainty from $t^\star$.
Screening is simply omitted for the IR safe terms because their contribution is 
comparatively small, as we shall discuss later.

Figure \ref{fig: eta HTL} also shows that, despite the aforementioned uncertainties, $\eta_\LO$
could be compatible with $\eta/s \lsim 0.5$. 
The rigorous bound $s > 4T^3$ of the interacting entropy for $T>1.2 T_c$ is known from lattice calculations \cite{Giusti2017}.
Thus the resummed LO viscosity may actually provide a prudent basis for extrapolation, 
rather than grossly overestimating $\eta/s$, it would actually give $\eta/s < 1/(4\pi)$ for large enough $\alpha$.
That brings us back to the central issue of coupling renormalisation at finite-$T$, \ie\ what is $\alpha$?

\subsection{Running coupling \label{sec 3B}}

For QCD there is no preferred value for $\alpha$ (such as $\alpha_{\rm em}=\frac{1}{137}$ in QED
for all practical purposes).
Loop corrections include both vacuum effects and (finite) thermal fluctuations, 
of which the latter survive the classical `$\hbar \to 0$' limit.
To then self-consistently define the running coupling,
even at the LL level, one must retain the first term in \eq{Pi}.
In Coulomb gauge, this is easily demonstrated because of the simple Ward identities 
already mentioned.

Typical loop integrals in quantum field theory give infinities which require renormalisation: 
connecting renormalised parameters with observables at the renormalisation scale.
A textbook result \cite{ItzyksonZuber}, handled through dimensional regularisation with the auxiliary scale\footnote{
  In the familiar $\overline{\rm MS}$ scheme, to absorb the universal constants.} 
$L$, isolates the divergent term
\be
\Pi_{\mu\nu} \Big|_{T=0} 
=
\alpha \beta_0 \big( \, Q^2 g_{\mu\nu} - Q_\mu Q_\nu \, \big)
\Big[\  \frac{1}{\epsilon} + \log \frac{-Q^2}{L^2} \ \Big] \ ;
\quad \epsilon \to 0 \, .
\label{Pi0}
\ee
Here $\beta_0 = (33 - 2 \nf)/(12 \pi)$ for QCD, where a positive value $\beta_0>0$ signals 
{\em antiscreening} of colour charges.
In vacuum, the transverse and longitudinal projections  coincide and are equal to
$\Pi^{\rm vac} (Q) \equiv\, \alpha \beta_0 [\epsilon^{-1}+\log(-Q^2/L^2)]Q^2 $\, .

\begin{figure}[tb]
  \centerline{\includegraphics[scale=1.]{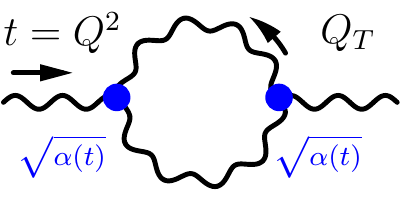}}
  \caption{
    Example of a 1-loop contribution to the gluon self-energy.
    Regularising the vacuum contribution, cf. Eq.~\eq{Pi0}, 
    we illustrate that the external momentum $t$ is the correct
    scale for the coupling in thermal mass $\widetilde\Pi(Q) \sim \alpha(t=Q^2)T^2$ --
    not the typical loop momentum $Q_T \simeq 2\pi T$ often presumed.
  }
  \label{fig: Pi}
\end{figure}

This brings us to the crux of our argument, which parallels the standard scenario at $T=0$ :
Since the scale $L$ was auxiliary, we use it as the scale with which we fix parameters in the theory.
Consider an experiment at the scale $\hat t=-L^2$, thus producing a finite value for $(\alpha^{-1} - \epsilon^{-1})$ which {\em defines}
an effective coupling $\hat \alpha^{-1}$.
(I.e. the renormalised coupling  at the scale $\hat t$.)
This is sufficient to make predictions at $Q^2\neq \hat t$ and leads directly to the 1-loop {\em running coupling} formula,
\bea
\alpha (Q^2) &=&
\big[ \beta_0 \log (-Q^2 / \Lambda^2 ) \big]^{-1} \, ,
\label{running coupling}
\eea
where the scale $\Lambda$ is determined by $L$ and the accompanying $\hat \alpha$\,.
Dressing an internal propagator $D_{\mu\nu}$, 
leads to the {\em renormalised} amplitudes [for the replacement rule \eq{IR replacement}]
\bea
\alpha \Delta (Q) &\to& 
\frac{\alpha}{Q^2 - \big(\, \Pi^{\rm vac}+ \wv\Pi\, \big) }
\ = \
\Big[\ \frac{Q^2}{\alpha(Q^2)} - 4\pi T^2 \, \phi (\omega, \bm q)\ \Big]^{-1} \, .
\label{M thermal}
\eea
For large enough $-Q^2$ (due to asymptotic freedom; the screening term is insignificant in the denominator),
we recover the familiar result, \eg\  ${\cal M}^{\rm vac} \sim \alpha(t)/t$ for a $t$-channel process like Fig.~\ref{1234}.
In this sense, the relevant scale in $\alpha(\cdot)$ is dictated by the process \cite{Cutler1978}.
For $T>0$, the running coupling also applies to thermal screening $\wv\Pi(Q) \sim \alpha(Q^2)T^2$, see Fig.~\ref{fig: Pi}.
Screening is then a genuine `perturbation' for $|t|^{1/2} \gg \Lambda$ because $\alpha$ is small, 
although (as mentioned) HTL propagators are no longer valid.
Notwithstanding soft momenta $|Q^2| \simeq \Lambda^2$, where screening is crucial, 
the ratio $Q^2/\alpha(Q^2)$ is then small by assumption\footnote{
  This, despite the Landau pole at $Q^2 = \Lambda^2$ in $\alpha(\cdot)$. 
  We shall prudently avoid these (meaningless) large $\alpha$ values 
  by using \eq{alpha eff}.
}.
Thus where pQCD becomes dubious, the matrix element saturates at
some finite value, rather than giving an obviously unphysical result.
At `strong coupling' the totally screened cross section actually leads to 
a {\em minimum} bound on $\eta$.
On these grounds, an extrapolation of $\sigma_{\rm tr}$ using Eq.~\eq{M thermal}
may actually yield reasonable estimates, e.\,g. for $\eta$, since screening protects $|t|^{1/2}
\to \Lambda$.

\bigskip
Strictly speaking, the `right' value of $\alpha(\cdot)$ should be
established by calculating all ${\cal O}(\alpha)$ corrections.
Evidently we have not done this.
Instead we advocate resumming a subset of these:
Those which come with large logarithms $\alpha (Q^2) \log (Q^2/t)$, and can be
eliminated by choosing $Q^2 = t$. 
We cannot say whether the uncalculated corrections will be large or small (see discussion
in Sec.~\ref{sec:sigTR}), but it seems legitimate to ask what effect it has.
Because it is based on adding vacuum corrections to the (already) minimal requirement for screening,
we speculate that it is also important for extrapolating  the LO result.

\subsection{Treatment of `subleading' terms}

The dominant contribution to $\eta$ comes from terms in $\d \sigma^{\rm tree}/\d t$
that resemble \eq{M thermal}, and also give the LL result in a parametric calculation.
Having found large (but not unreasonable) sensitivity to the limited scope of the HTL functions
(see Fig.~\ref{fig: eta HTL}), we can afford to make some simplifications in subleading terms.

\begin{figure}[thb]
  \includegraphics[scale=\FigScale]{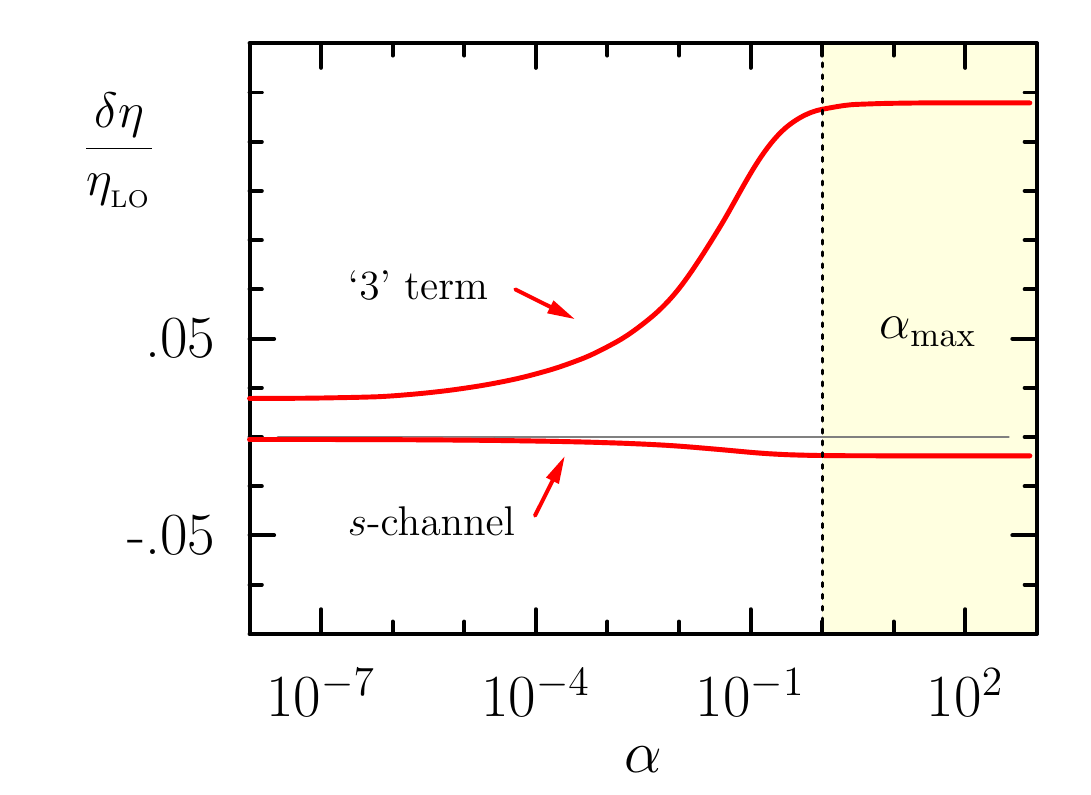}
  \caption{ 
  Relative contribution from subleading terms in $\d \sigma^{\rm tree}/\d t$, 
  which do not require screening in Eq.~\eq{glue-glue} (hence no sensitivity to $t^\star$).
  Here $\delta \eta = \eta_\LO - \bar\eta$ where $\bar\eta$
  is without certain contributions (labelled by arrows) to $\d \sigma^{\rm tree}/\d t$.
   Then $\delta \eta/\eta_\LO$ is a  
  `fraction' of the Red curve in Fig.~\ref{fig: eta HTL}.
  The $s$-channel contribution is negative, but small.
  A maximal coupling value is from \eq{alpha eff}.
}
  \label{fig: relative}
\end{figure}

That is the reason we have altogether dropped inelastic processes, which affect $\eta_\LO$ by
merely a few percent \cite{Arnold2003f}.
Similarly, the terms in $\d\sigma^{\rm tree}/\d t$ that do not need to be screened
give a numerically minor contribution (screening is entirely omitted for them); 
see Fig.~\ref{fig: relative}.
To correctly incorporate screening into the sub-dominant and $s$-channel process would require
dressing the individual amplitudes ${\cal M}_i$, rather than selectively mending the total cross section
$\d \sigma / \d t \sim | \sum {\cal M}_i |^2$.
Hence we cannot preclude $|{\cal M}^2|<0$ in
the numerical evaluation  of ${\cal C}[\chi]$, but it seems to carry barely any influence on its value.
(Figure~\ref{fig: relative} shows why; the offending terms are subleading.)

Equation~\eq{running coupling} applies to the  space-like exchanges
in $\d\sigma^{\rm tree}/\d t$ that require thermal screening.
Although it would be tempting to simply evaluate $\alpha(Q^2)$ at
the virtuality of the intermediate state $Q^2=\{s,t,u\}$, that would in any case
not help for the inelastic scatterings.
We motivated a running coupling only for the leading IR terms.
For the rest, let us simply take $Q^2 = (stu)^{1/3}$, and use
the continuation of \eq{running coupling}, as advocated in
\cite{Shirkov1997a}
\bea
\alpha_{\rm eff} (Q^2) &=& \frac{1}{\pi \beta_0} A\big(Q^2/\Lambda^2\big) \, ,
\label{alpha eff}
\eea
where the (analytic) function $A$ is defined by
\bean
& & A(y) = \l\{ 
\begin{array}{l}
  \D \frac{\pi}{\log( -y)} + \frac{\pi}{1+y} \\[.4cm]
  \D \frac{\pi}{2} - \arctan \Big(\, \frac1\pi\log \, y\, \Big)
\end{array}
\r. \quad \text{for} \ y \lg 0 \, . 
\eean
Despite intrinsic difficulties with QCD in
the far infrared, perturbation theory can give semi-quantitative results
\cite{Dokshitzer2002}.
This model expression \eq{alpha eff} has a `universal' limiting 
value at $Q^2 \to 0$, (from above or below),
that imposes a maximal value $\alpha (\cdot) \leq \alpha_{\rm max} = 1/\beta_0$.
Numerically $\alpha_{\rm max} = \{ 1.1, 1.3 \}$ for $\nf = 0$ and $3$ .
and having larger values of $\alpha$ (\ie\ from using the na\"{i}ve one-loop
formula with $\alpha_{\rm max} \gsim 1$ imposed by hand)
does not markedly change the results, see Fig.~\ref{fig: relative}.
The reason for this was already discussed around Eq.~\ref{M thermal}.

%% file: section-4.tex
\section{Results}
\label{sec:results}

\begin{figure}[hb]
  \includegraphics[scale=\FigScale]{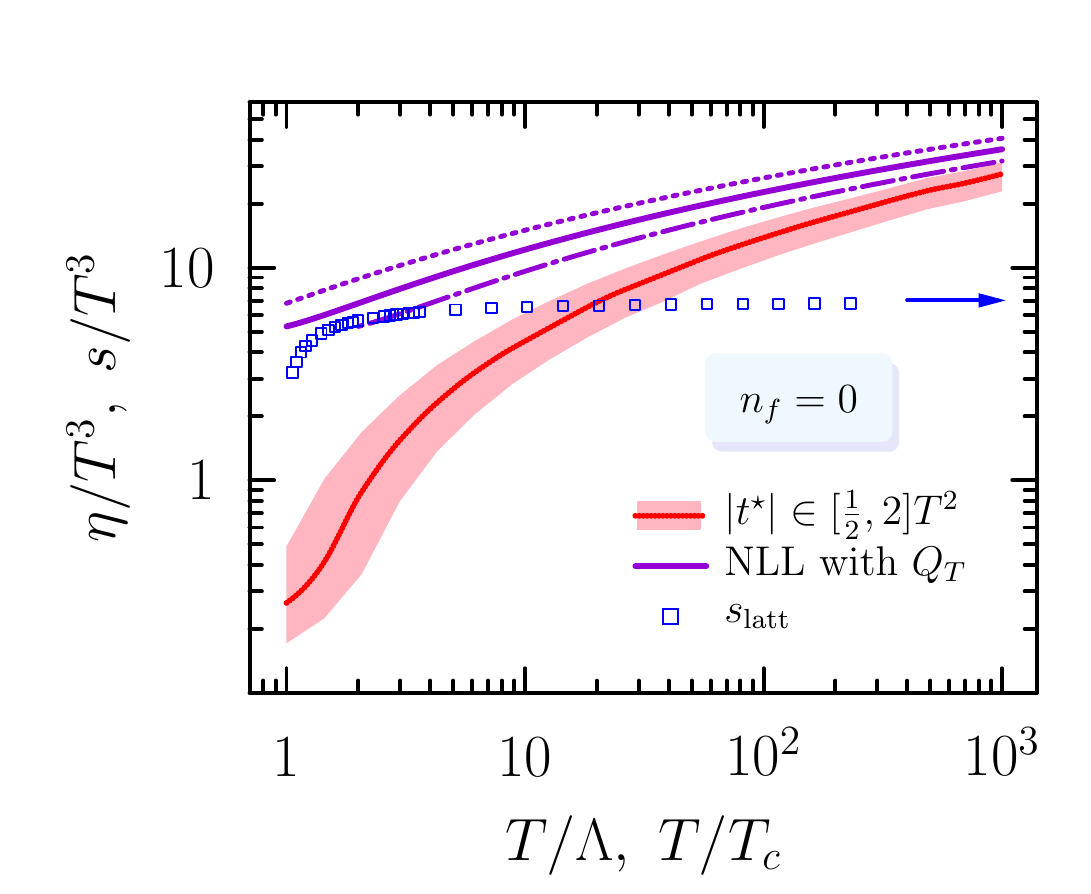}
  \caption{
    Temperature dependence of the viscosity in the quenched limit
    ($\nf=0$).
    The purple curves (solid, dashed and dotted)
    represents the NLL result for $\eta$ (see Fig.~\ref{fig: eta(alpha)}),
    and with a coupling at the 
    thermal scale $Q_T = 4 \pi T$; dotted, $2\pi T$; solid and $\pi T$; dash-dotted.
    The solid (red) line, surrounded by a light band, is our 1-loop resummation
    scheme (see text for
    details) with a `canonical' choice of screening only where $|t| < -t^\star \sim T^2$.
    The blue squares are the lattice results \cite{Giusti2017} and the arrow points to the 
    Stefan-Boltzmann limit.
  }
  \label{fig: eta(T)}
\end{figure}

Given then, is $\eta/T^3$ as a function of $T/\Lambda$
via the scales that go into $\alpha(\cdot)\,$.
Figure \ref{fig: eta(T)} shows the resulting temperature dependence for 
$\eta_{_{\rm LO}}/T^3$,
determined using Eq.~\ref{alpha eff} with an appropriate scale
 in the renormalised HTL propagators. 
 [The details are in Sec.~\ref{sec:renorm}, and actual $Q^2$-values are intergration variables for \eq{baym}].
Also shown in Fig.~\ref{fig: eta(T)} is the NLL result \eq{eta NLL}, with a running coupling
$\alpha(Q_T^2)$ where $Q_T = \xi \cdot 2\pi T$ for $\xi \in [\frac12,2]$.
Evidently the two curves approach one another for $T \gg \Lambda$,
still differing by a factor $\approx 2$ at $T/\Lambda = 10^3$,
for which the na\"{i}ve coupling is $\alpha (Q_T) = 0.1\,$.
Note in particular, that $\alpha (2\pi T_c/\Lambda) \approx 0.3$ -- just
about $\u{\alpha}$ where \eq{eta NLL} breaks down.
Changing $\xi$ is akin to a trivial rescaling of $\Lambda$ (since only the ratio 
$Q_T/\Lambda$ is important for $\eta_{_{\rm NLL}}/T^3$), and thus
merely shifts the corresponding curve in Fig.~\ref{fig: eta(T)} to the left or right.

Plotting the interacting entropy $s(T)/T^3$, obtained now
to high precision for $T\leq 300 T_c$ \cite{Giusti2017}, on
the same axis illustrates $\eta = s$ only at moderately large temperatures $T \simeq 20 T_c$ (assuming $\Lambda=T_c$).
The entropy (being a measure of the available degrees of freedom) increases rapidly at 
at the transition, and is within 5\% of the Stefan-Boltzmann limit after
$T\gsim2T_c$.
The shear viscosity (divided by $T^3$) increases more gradually above $T_c$, due to 
asymptotic freedom. [The mean free path becomes small, due to $\lambda \propto T^{-1}$, but
but also depends weakly on $\alpha$, see \eq{eq1}.]
However for physically relevant temperatures,
just above the crossover (say $T<4 T_c$),
the renormalised prediction is an order of magnitude smaller than the
na\"{i}ve application of the NLL result.
And while there is no guarantee that pQCD (or indeed kinetic theory) is applicable here, 
it is just what one would expect from a reasonable extension of the curve at $T\simeq 20T_c$ (where $\eta=s$).
We shall speculate on this extrapolation after first 
presenting the subsequent ratio $\eta/s$, for $T\gsim T_c$.

\subsection{Estimation of $\bm \eta \bm/ \bm s$}

The interacting entropy is now well established in the published lattice QCD
 data (available for $\nf=\{0,3\}$) \cite{Giusti2017,Bazavov2014f}, 
and we shall use it to normalise our perturbative result for $\eta$.
Adjusting the quantity $\ell=\Lambda/T_c$ in the ratio
$$
\eta \l( T/ \bm( \ell\,T_c \bm) \r) \big/  s_{\rm latt} \l( T/T_c \r) \, ,
$$
overlays the units for the abscissa in  in Fig.~\ref{fig: eta(T)}.

\begin{figure}[bt]
  \includegraphics[scale=\FigScale]{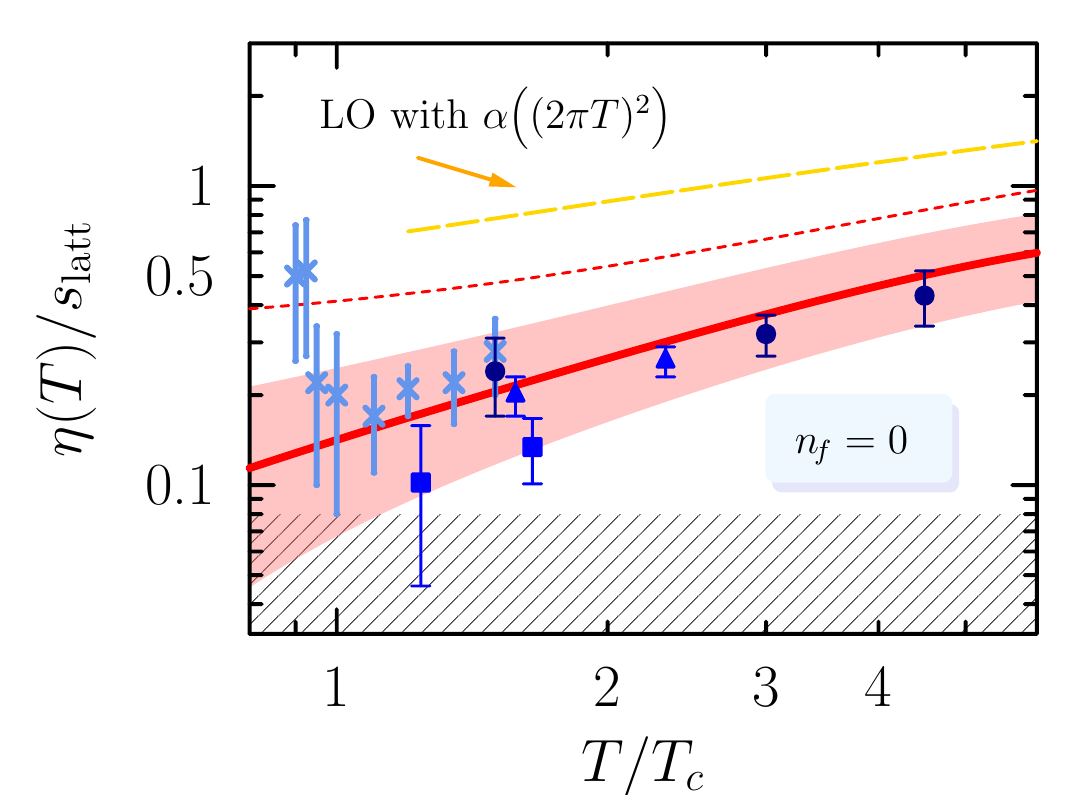}
  \caption{
    A renormalised prediction for $\eta/s$, as a function of $T/T_c$, 
    in pure gauge QCD (assuming $T_c/\Lambda=1.26$).
    The lattice data are from Refs. \cite{Meyer2007f}; $\Box$,
    \cite{Meyer2009e}; $\triangle$, \cite{Mages2015}; $\circ$ and 
    \cite{Astrakhantsev2017}; $\bm\times$\,.
    The dashed curve (orange) is the LO result from \cite{Arnold2003f},
    with a running coupling at the `hard' thermal scale $Q_T = 2\pi T$.
    Hatching indicates the region below the
    conjectured lower limit $1/(4\pi)$.
  } 
  \label{fig: eta/s quenched}
\end{figure}

\bigskip
For $\nf = 0$, there are several evaluations of $\eta_{\rm \,latt}$ 
\cite{Meyer2007f, Meyer2009e, Mages2015, Astrakhantsev2017}.
(Caveats, mentioned in Sec.~\ref{sec:intro}\,, should be reiterated here:
A dynamical quantity is difficult to extract by means of {\em equilibrium} lattice gauge theory.)
Nevertheless, if we confront the LO calculations with $\eta_{\rm\,latt}$ at the simulated temperatures and 
use\footnote{
This differs from \cite{Jackson2017a}, where we simply set $\ell \to 1$ because $\Lambda = {\cal O}(T_c)$. 
Our results are not too sensitive to its value, as to be expected.}
$\ell = 1/1.26$ \cite{Borsanyi2012e}, we find that $\eta_\LO(T)$ is quite compatible.
Figure \ref{fig: eta/s quenched} indeed reveals that our calculation corroborates the lattice data,
and can explain $\eta/s \approx 0.2$ at $T < 2T_c\,$.
Modifying the self energy in the screened propagators via $t^\star \in -[\frac12,2]T^2$
gives a family of curves that cover the estimated uncertainty from $\eta_{\rm\,latt}$.
Setting $t^\star/T^2 = -\infty$, \ie\ using HTL functions for all $\omega$ and $|\bm q|$,
gives $\eta$ about a factor of two larger.
This is to be expected, based on the similar sensitivity for the toy model in Sec.~\ref{sec:sigTR} 
[see eq.~\eq{mu2 with t^star}].

Using the coupling at an argument $Q_T=2\pi T$ would instead
yield $\eta_{_{\mathrm{LO}}}\bm(\alpha(Q_T)\bm)/s \gsim 0.7$, which still 
does not explain $\eta/s < 0.5$.
Adjusting (via $\xi$) the scale $Q_T=\xi\cdot 2\pi T$, to have 
$\eta_\LO(\alpha)/s_{\rm latt} < 0.5$ at $T=1.2T_c$ would require $\alpha\gsim 0.4$ 
(see Fig.~\ref{fig: eta HTL}).
It then seems difficult to justify the 
resulting $\xi \lsim 0.3 \Lambda/T_c$ from using \eq{alpha eff}\footnote{
  The unmodified coupling \ref{running coupling} would require $\xi \lsim 0.5 \Lambda/T_c$ instead.}.
Moreover, the corresponding curve would give the wrong slope for $\eta/s$ as a function of $T$.
Confinement sets in for $T<T_c\,$, where the unmistakable increase in $\eta/s$
is due to both a reduced entropy and larger viscosity:
$\eta_{_{\rm glueball}} \sim (\Lambda/T)^{5/2}$ \cite{Hosoya1985a}.

All our discussion trivially extends to the physical case, $\nf=3$, for which
it ought to viewed in the context of RHIC and LHC programmes.
There it has been established that the experimental data 
cannot be reproduced with a value $\eta/s > 0.5$,
by viscous hydrodynamic simulations \cite{Teaney2003a}.
$\eta/s$ is one input to these codes, whose value is tuned to cover the hadronic and QGP 
phase of the evolution near $T_c\,$.
Studies first sought a feasible range for $\eta/s$
(to reproduce experiments):
\be
\begin{tabular}{c || c | c | c | c | c | c}
  Ref. & \cite{Aamodt2011} \& \cite{Adare2011} & \cite{Luzum2008a} & \cite{Song2011e} & \cite{Gavin2006a} \\
  \hline
  \multirow{2}{*}{$\D \frac{\eta}s \gl$} & $0.11$ &  & $.08$  & $0.08$ \\
                                         & $0.16$ & $0.28$ & $0.20$ & $0.3$
\end{tabular} 
\label{min eta/s}
\ee
Recently, attempts have been made to constrain the temperature dependence of $\eta/s$, 
which is expected to be minimal at the crossover temperature \cite{Niemi2015qia, Denicol2016, Bernhard2016}.
These approaches are capable of eliminating certain {\em models} for $\eta(T)$;
which also depends on $T<T_c$ (the hadronic phase of evolution).
Simple parametrisations of $\eta/s$ are linear in $T$ 
with different slopes for $T\gsim \u T$ (where $\eta/s$ is minimal at $\u T\approx T_c\,$).
Exploring minimal values from \eq{min eta/s} offers some restrictions on the two slopes --
although it remains difficult to discern the behaviour for $T>\u T$\,.

\begin{figure}[htb]
  \includegraphics[scale=\FigScale]{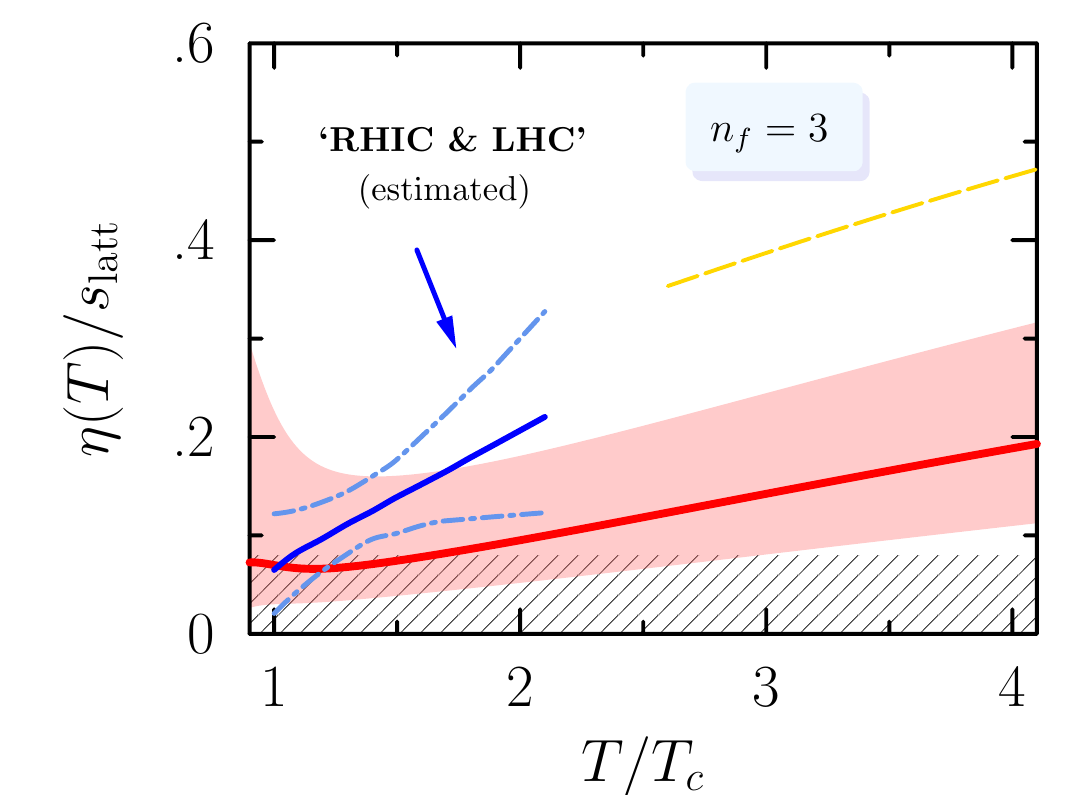}
  \caption{
    The normalised viscosity for $\nf=3$ flavours as a function of $T/T_c$ 
    (here $\Lambda/T_c = 2.1$ \cite{Aoki2009n}\,).
    Our estimate for the sensitivity to higher order terms gives the
    band around the solid red curve (see Fig.~\ref{fig: eta HTL}).
    Blue curves illustrate the `most likely' dependence
    from a Bayesian analysis using 
    hydrodynamical simulations \cite{Bernhard2016}.
    (Mean; solid blue and 95\%-confidence; dash-dots.)
    Hatching indicates $\eta/s \leq 1/(4\pi)$.
  }
  \label{fig: eta/s nf=3}
\end{figure}

In Fig.~\ref{fig: eta/s nf=3} we show the viscosity to entropy density
in the physical case, where we have set $\ell = 1/0.48$ \cite{Aoki2009n}.
Our results are on the lower end of the parametrisations explored in hydrodynamics
and predicts a milder increase with $T/T_c\,$.
Apparently there is hardly any difference with the quenched results;
the increased interaction rate is compensated by the density.
There does seems to be more sensitivity to the HTL restriction from $t^\star$,
which brings a factor $\approx 2$ uncertainty.
For $T<3T_c$, it seems quite possible to violate the lower bound $1/(4\pi)$
-- or not, depending on $t^\star$.
As speculation, the smooth crossover in $s(T)$ (compared to $\nf=0$)
may cause the minimum in $\eta/s$ to be shifted slightly; $\u T \approx 1.5\,T_c$.
(The fact that $s_{\rm latt}$ deviates from $s_0$ by 30\% at $T=2T_c$ may
indicate a change in quasiparticle structure already, which we assumed not the case for $\eta$.)

The ratio of the LO viscosity with $\alpha(Q_T^2)$ versus our results is at the factor-of-2 level
(for $T=4T_c$, see Fig.~\ref{fig: eta/s nf=3}).
So extrapolating the fixed-coupling results with $Q_T$ down to $T\sim 1.5T_c$, would give $\eta/s \approx 0.3$
(by eye) which is possible.
However, the renormalised viscosity reaches this value already at $T\approx 5T_c$.

\subsection{Kinetic Theory \label{sec 3C}}

We now pause to confront the question of whether kinetic theory is applicable in 
such a `strong coupling' regime.
At asymptotically high $T$, there is a clear ordering
between the inter-particle distance $\bar r \sim T^{-1}$, the Debye screening length
$m^{-1}_D \sim (\sqrt{\alpha}\,T)^{-1}$ and the (large or small-angle scattering) 
mean free path $\lambda \sim (\alpha^j\, T)^{-1}$ where $j=\{1,2\}$ \cite{Arnold2003a}.
So
$$ \lambda \gg m_D^{-1} \gg \bar r $$ 
and it is permissible to neglect multi-particle correlations
and treat the individual {\em binary} collisions as instantaneous.
But the scales become comparable for
achievable temperatures $T \approx T_c\,$.

For example; the Debye mass (computable on the lattice for $\nf=0$)
is largest when $T= 1.2 T_c$ but
still satisfies $m_D^{-1} \geq \frac13 T^{-1}$ \cite{Peshier2006}.
The interparticle distance 
$\bar r \approx \frac12 n^{-1/3}$ follows\footnote{
  There could be other definitions of $\bar r$, \ie\
  setting it equal to $n^{-1/3}$ would amount to `dense packing'.
  We consider the `nearest-neighbor' definition discussed in many textbooks --
  hence a factor $\frac12$ \cite{Reif1964}.
}
from the number density $n=(16+\frac{21}{2}\nf)\frac{\zeta(3)}{\pi^2} T^3$.
Hence the effective range of interactions is approximately equal to $\bar r$,
which puts the physical picture at its limit but does not clearly invalidate it.
Kinetic theory cannot be right at $T_c\,$, and thus we rather want to know where it actually 
breaks down.

\begin{figure}[hb]
  \includegraphics[scale=\FigScale]{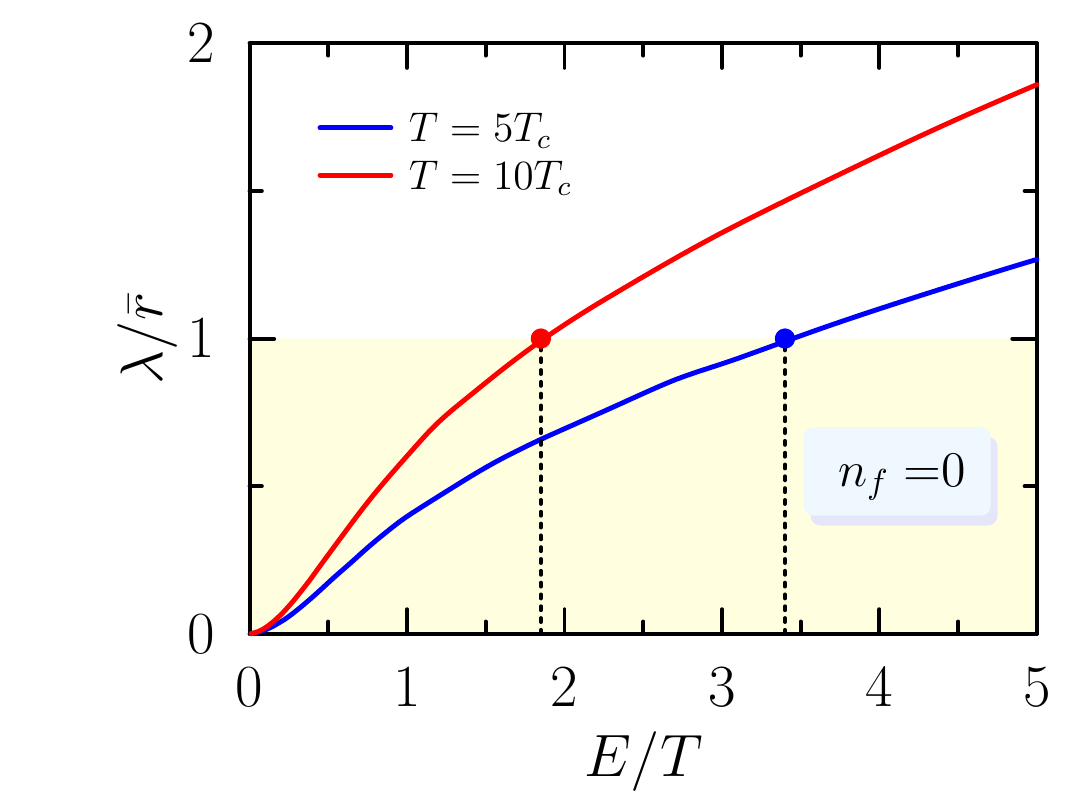}
  \caption{
    Mean free path $\lambda\,$, for a gluon of energy $E\,$, in units of the mean
    interparticle distance $\bar r\,$.
    Two temperatures are shown (for $\ell=1/1.26$ as in Fig.~\ref{fig: eta/s quenched}).
  }
  \label{fig: mfp}
\end{figure}

Until now we have simply used the transport equation \eq{C[f]} as a starting point,
and (with motivation) applied it where $\lambda$ may not be large (\ie\ strong coupling).
To justify this idea {\em a posteriori}, we show that the mean free path $\lambda$
comes to the order the inter-particle distance $\bar r$ at a only few times $T_c\,$.
To define $\lambda$, we shall appeal to the relaxation approximation
${\cal C}[f_1] = -\delta f_1/\lambda$ in the Boltzmann equation \eq{C[f]}.
After linearising, $\lambda(E_1)$ is expressed in terms of the collision
operator,
\be
\lambda^{-1} (E_1) =
\frac{1}{6E_1} \int \d \Gamma \,
\frac{tu}{s^2} |{\cal M}|^2 \frac{ f_2 \bar f_3 \bar f_4 }{\bar f_1} \, .
\label{absorption rate}
\ee
The transport factor $tu/s^2 = \frac14 (1-\cos^2\theta)$ comes from 
$\hat p_1^i \hat p_1^j \Delta^{ij}[\chi]\big|_{\chi=p^2}$ evaluated in the centre of momentum frame ($\theta$ is the scattering angle).
We may then calculate \eq{absorption rate} with the screened QCD matrix elements ${\cal M}$,
and using the running coupling as put forward in Sec.~\ref{sec 3B}.
Figure \ref{fig: mfp} shows the resulting $\lambda(E)$ in units of $\bar r$, at $T=\{ 5T_c,10T_c \}$.
(The variation from modifying $t^\star$ from Sec.~\ref{sec 3A} is on par with a 
factor of 2 in temperature.)

What fraction of particles is able to deliver its momentum across a path length $\lambda\geq \bar r$?
If $\lambda=\bar r$ at $E\to E^\star$ then this fraction is $ \int_{E^\star}^\infty \d E\, E^2 f(E)\big/ (2\pi^2 n)\, $.
For the two temperatures shown in Fig.~\ref{fig: mfp}, we find that $70\%$ ($T=10T_c$) and $40\%$ ($T=5T_c$) particles meet this
(albeit slightly arbitrary) requirement.
Note that $T=5T_c$ is roughly the highest temperature used in lattice calculations
of $\eta$ (see Fig.~\ref{fig: eta/s quenched}).
Using $\bar r$ to qualify kinetic theory is not a rigorous judgement,
but it at least demonstrates that we are at its limit
for relevant temperatures $T \to T_c\,$.

For $\nf=3$, quarks and gluons will have different mean free paths.
Because of the group factors in their interaction rate, $\lambda_g < \lambda_q\,$.
Unfortunately there are large uncertainties coming from $t^\star$ in \eq{absorption rate}, hence we
do not show a plot like Fig.~\ref{fig: mfp}.
But is seems that $\lambda_g \lsim \bar r \lsim \lambda_q$ for the bulk of particles 
at a few times $T_c\,$.

%% file: section-5.tex
\section{Summary}

In this paper we refute a widespread paradigm: That the inferred ratio of shear viscosity 
to entropy density of the QGP, $\eta/s \lsim 0.5$ near the deconfinement temperature $T_c$,
is a genuinely non-perturbative effect.
Contrary to this view, we find that the LO weak coupling result of Ref.~\cite{Arnold2003f} 
can explain {\em why} $\eta$ is so low.
Our main result is a new estimate for the temperature dependence of $\eta/s$,
see Figs.~\ref{fig: eta/s quenched} and \ref{fig: eta/s nf=3}. 
\bigskip

We addressed two problems with many existing estimates from perturbation theory:
\begin{enumerate}
  \item[1.] 
    Logarithmic accuracy is appropriate only in scenarios where the coupling  may
    be regarded as (asymptotically) weak. The increase of $\eta_{_{\rm NLL}}(\alpha)$
    for $\alpha > \alpha^\star$ is {\em unphysical}: we expect that the viscosity
    to decrease with the  coupling strength.
  \item[2.] 
    A common procedure in using a fixed coupling result, is to  {\em choose} the 
    value of $\alpha$ as the running
    coupling $\alpha(Q^2)$ at, or near, the lowest Matsubara energy.
    In fact, a scale-dependent coupling and the calculation scheme of LO
    approximations are closely related: both emerge from/take into
    account loop corrections to tree level-amplitudes.
\end{enumerate}
Together, these improvements enable us to meaningfully extrapolate 
$\eta (T)$ near the QGP phase transition, 
provided that binary scatterings form the principal source of energy and momentum
variation.

To see how the running may affect a coefficient like $\eta$, consider our earlier formula \eq{eq1} 
for the transport cross section.
Using \eq{running coupling} in \eq{eq1} to LL accuracy, we find\footnote{
A similar setting of scales, balanced between hard and soft modes, was 
found for the QCD collisional energy loss \cite{Peshier2006a}.}
\bea
\sigma_{\rm tr} (s)
\sim
\frac 1s
  \int_{-s}^{-\mu^2} \d t\, |t| 
  \, \frac{\alpha(t)^2}{t^2}
\ = \
\frac 1s
\alpha( \mu^2 ) \alpha (s) \log \Big(\, \frac{s}{\mu^2} \,\Big)\, .
\label{pocket}
\eea
While the overall structure is unchanged, cf.~\eq{eta/s parametrically}, 
 a substitution 
$\alpha^2 \to \alpha (\text{\sl hard})\alpha(\text{\sl soft})$ in going from
\eq{eta NLL}  to \eq{pocket}
reflects the relative importance of different scales.
Soft interactions are more probable due to an overall $\alpha(t)^2$ in $\d \sigma /\d t$,
but are also more screened by $\mu^2 \sim \alpha(t)T^2$.

Resumming the vacuum self-energy $\Pi^{\rm vac}$ in \eq{IR replacement} 
is `optional', only thermal corrections
{\em must} be taken into account (running in $\alpha$ is formally higher order).
However, we may incorporate them so that vacuum and thermal parts are treated on an equal footing. 
And if we do, the benefit is to specify $\alpha$ (even at the LL level).
In pursuing this line of reasoning, we have found that the scales in the running coupling
could have a substantial effect on $\eta$.

%% file: appendix-1.tex
\def\wv#1{\widetilde{#1}}
\section{$\bm{P(s)}$ distribution\label{A5}}

A valuable clue as to the role of the `hard' scale $s \sim T^2$, appearing in
Eq.~\eq{m1}, is provided by the thermal weight $P[\chi]$, emerging from \eq{baym},
whose properties we now explore.

Firstly, the condition for normalisation, \ie\ $\int \d s P(s) = 2b$, 
is convenient for \eq{eta with P(s)}  but should be set by $\chi$.
After using symmetry in the integration variables $p_{1,2}$ of 
\eq{Pdef}, we may complete all but one integral to find
(for notational convenience we put $T \to 1$ here)
\bean
  \int \d s P(s) &=&
  \frac23 (2\pi)^{-2}
  \int_0^\infty \d p f \bar f
  \Big[ \, \chi^2 + \frac{p^2}{6} (\chi^\prime)^2 \, \Big] \, .
\eean
Requiring that this equals $2b$ imposes a restriction
on the function $\chi$, that needs to be satisfied in addition
to the constraint \eq{B}.
Taken together, they imply $\chi$ must satisfy 
\bea
\chi^{\prime\prime} + 
\Big( \, \frac2p - 1 - 2f \, \Big) \chi^\prime
- \frac{6}{p^2} \chi + A p &=& 0 \, .
\label{DE}
\eea
$A$ is a Lagrange multiplier to set the norm of $\chi$ 
(it cannot be zero).
This relation was also found in Ref.~\cite{Heiselberg1994b} in a different manner,
directly evaluating \eq{Q} to LL accuracy.
Because the solution to the Boltzmann equation 
$f_\star (p) =f [1+ \chi \bar f (\hat p_i \hat p_j - \frac13 \delta_{ij})\nabla_i u_j]$
is positive and integrable, the function $\chi$ cannot grow exponentially fast.
Suppose that $\chi(p)$ is normalised, in such a way that 
$\chi \to p^\nu$ for large $p$.
Substituting into \eq{DE} with $f\approx0$ gives $\nu=2$ and $A=2$.
(Rescaling the asymptotic behaviour by some constant will just change $A$ by the same factor.)
\bea
\chi (p) &\sim&
p^2 + O(p) \, ;\quad p\gg T\, .
\label{chi for large p}
\eea
In Ref.~\cite{Baym1990}, the optimal solution
to \eq{etaVAR} of the same `single power' form 
for {\em all} $p$ gave $\nu\approx 2.104$\,.

For small arguments, $f(p) \simeq \frac{1}{p} - \frac12 + \frac{1}{12}p + \ldots$
in \eq{DE} and the first two terms in brackets thus cancel.
The resulting asymptotic solution 
(contrary to the claim of \cite{Heiselberg1994b}) is
\bea
\chi(p) &\sim&
p^3 \Big(\, \bar c - \frac{2}{5}\log p \, \Big)
+ {\cal O}(p^5) \, ; \quad p\ll T \, ,
\label{chi for small p}
\eea
with $\bar c \approx 0.62$
(this integration constant is set by \eq{chi for large p}, for which we have only a numerical value).
Knowing the precise solution $\chi^\star$ to \eq{DE} only improves 
the estimate for $\eta$ in \eq{baym} by about $0.5$\%. 
The large-$p$ behaviour $\chi\sim p^2$ turns out
to be more important than \eq{chi for small p},
which is to be expected for a transport quantity.
We thus explore below, the function $P[\chi](s)$ at
the approximate solution $\chi \to p^2$.

\begin{figure}[hbt]
  \includegraphics[scale=\FigScale]{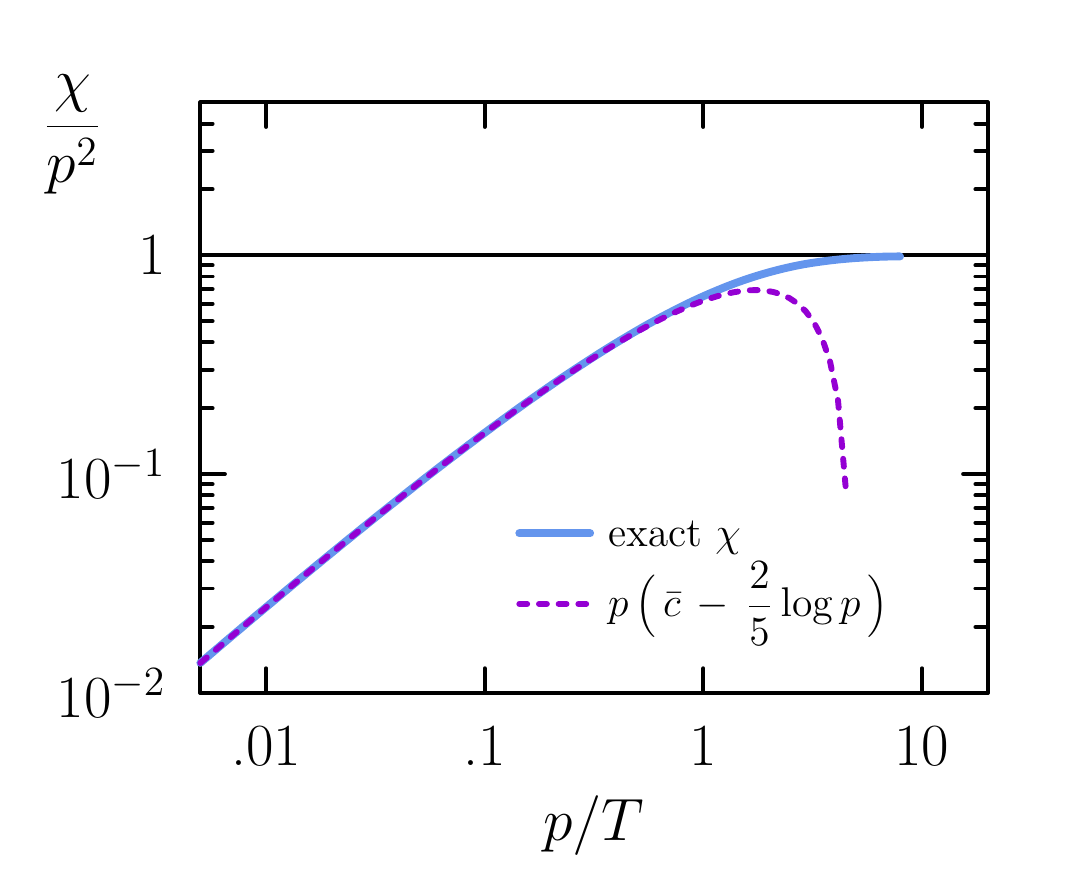}
  \caption{ 
  The function $\chi/p^2$ as determined by Eq.~\eq{DE}. 
  Also shown is the limit $\chi \sim p^3 \log 1/p$
  for $p \ll T$ (purple dotted line).
  $\chi\approx p^2$ appears to be an excellent approximation
  for $p\gsim T$.
  }
\label{fig:chi}
\end{figure}

\subsection*{Single-function Ansatz}

For a given collisional parameter $s$, using \eq{Pdef} with $\chi(p)=p^2$,
we define
\bea
{ P}(s) 
&=& 
 \frac{1}{(2\pi)^4}
 \int \d E_1 \d E_1 \,
 f_1 \bar{f}_1
 f_2 \bar{f}_2 
 \big[ \, 
  \frac{16}3 E_1 E_2 \l( E_1 - E_2 \r)^2 
  + \frac23 s \l( E_1^2+E_2^2  + 4E_1E_2 \r) + s^2 
 \,  \big] \, .\nonumber\\
\label{160211:eq1}
\eea
The integrals are over positive energies such that  $E_1 E_2 \geq s/4$\,.
We do not have a simple closed form expression for $P(s)$ beyond a 1-dimensional
integral over polylog functions, but we can state its large/small-$s$ behaviour 

For large $s$, the thermal distributions can be replaced by classical Maxwell-Boltzmann
distributions, i.\,e.\ $f_1 \to \exp (-E_1)$.
We then obtain
\bean
P_{\rm cl.}(s) &=& \frac{s}{15\cdot 2^{12}\cdot \pi^4 }  
\big[\, 
  (40+3s)\sqrt{s} K_3 \big(\sqrt{s}\big) - 6 s K_4 \big(\sqrt{s}\big) \,\big] \, ,
\eean
where $K_{3,4}$ are modified Bessel functions of the second kind.

\begin{figure}[hbt]
  \includegraphics[scale=\FigScale]{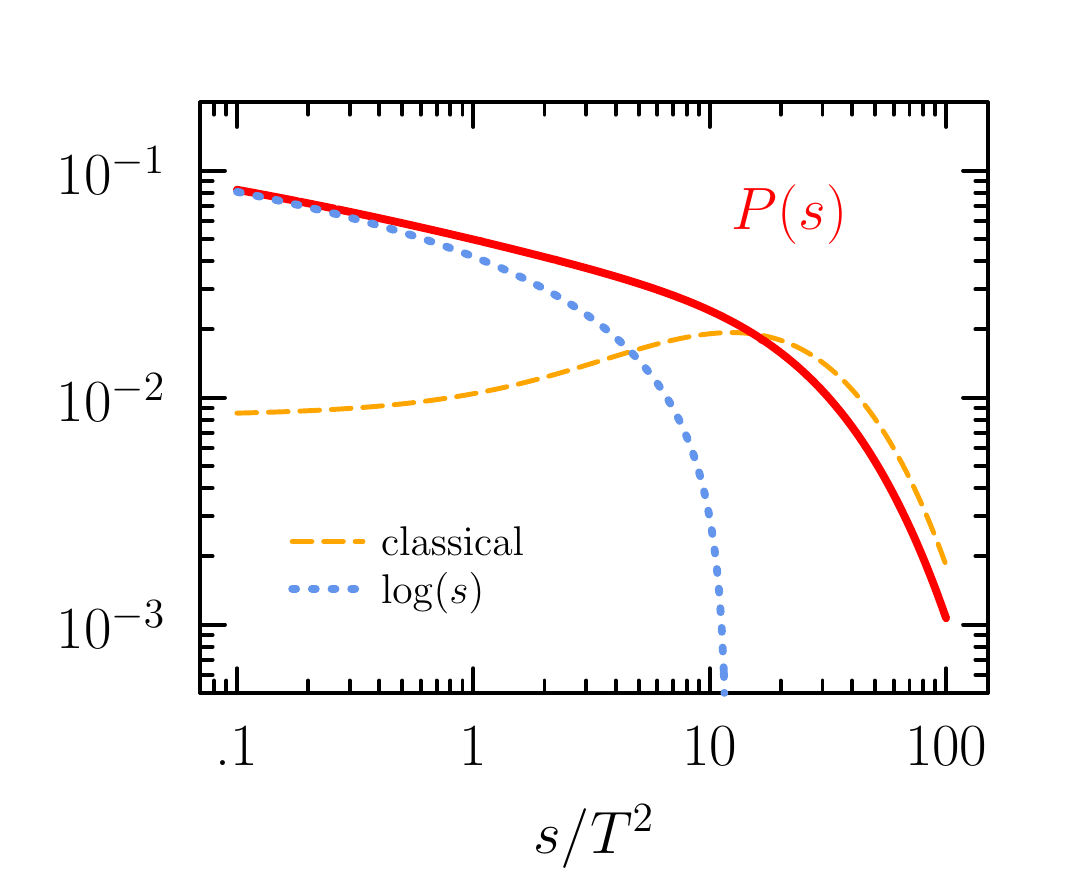}
\caption{ 
  The (normalised) probability, \eq{160211:eq1} divided by $M_0$\,.
  The small-$s$ expansion is
  shown for comparison (dotted blue line), and implies that the slope of $P$ at $s=0$
  is infinite. The counterpart function for classical Maxwell-Boltzmann particles
  is also shown (orange dashed line).
}
\label{fig:P(s)}
\end{figure}

The first few terms in an asymptotic series for $P(s)$ is given by
\bean
P(s) &\sim&
\frac{32}{27} \, \Big\{ \,
  54 \zeta^\prime (3) - \pi^4 +
  18 \big(\,  10 - 3 \gamma - 3 \log \frac{s}{4}\, \big)
\Big\} \\
& &  \
+\frac{2s}{9}\, \Big\{ \, 
66 - 12 \gamma^2 + \pi^2 - 54 \log \frac{s}{4} + 6 \log^2 \frac{s}{4} - 24 \gamma_1
\, \Big\}
+ {\cal O}(s^3) \, ,
\eean
where $\gamma_1$ is the first Stieltjies constant (the Euler-Mascheroni constant
$\gamma$ is $\gamma_0$).
The qualitative feature we needed for the argument in Sec.~\ref{sec:sigTR} 
is indeed confirmed;
$P(0) \neq 0$ (see Fig.~\ref{fig:P(s)}), in fact there is an 
(integrable) log-divergence at $s\to 0$\,.

The `moments' $M_n = \int_0^\infty \d s \, s^n P(s)$, are explicitly 
\bea
M_n 
&=&
\frac{ 4^{1+n} }{3\pi^4} \Gamma (n+1)  \Gamma (n+4)  \label{moments} \\
& & \qquad \times \Big[
  \zeta (n+4) \zeta (n+2) (n+4) (5+3n) 
  + \zeta (n+3)^2 n (7+3n) 
\Big] \, .
\nonumber
\eea
These moments grow extremely rapidly; $\log M_n \sim n \log n!$ as $n\to \infty$\,.
 Once rescaled, $P(s)/M_0$ can be interpreted as a (normalised) probability
 distribution which then gives
\bea
\l< s \r> = M_1/M_0 \simeq 27.7 T^2 \, ,\\
\l< \log (s) \r> = \pd_\epsilon M_\epsilon /M_0 \big|_{\epsilon\to0} 
\simeq 0.381 \, ,
\label{moments of P(s)}
\eea
required to specify the parameter $\kappa$ for the effective model in Section
\ref{sec 2A}\,.

The general formula for the moments \eq{moments} allows for analytic continuation to
 complex $n$, inheriting properties from the $\Gamma$- and $\zeta$-functions. Isolated
poles occur at $n=-1$ (double), $n=-2$ (triple) and double poles for all $n\leq -4$.
This reaffirms the fact that `negative' moments are divergent. 

%% file: appendix-2.tex
\section{Soft \& hard contributions\label{A4}}

Particle production rates and transport coefficients require kinematic integral
convolutions with matrix elements.
In a thermal setting, particle propagators are modified by their interaction with the
medium, a feature already stressed in Sec.~\ref{sec:sigTR}. The general form of this
screening is a $T^2$ multiplied by a dimensionless function of (separately) the
momentum and frequency. The HTL approximation \cite{Braaten1990c} affords an analytic
representation depending only on $z=\omega/|\bm q|$. While usually derived under the assumption that
both $\omega, |\bm q| \ll T$, the approximation is in fact valid for $|\omega^2 - q^2| \lsim T^2$  
\cite{Peshier1998}.

As mentioned in the body of the text (Sec.~\ref{sec 3A}), we opt for a covariant separation of phase
space with $t^\star$ as the cut-off parameter.
The only difference with \eq{mu2 with t^star} comes from the HTL self-energies.
Using the Born cross section where $-s <t < t^\star$, for the $t$-channel contribution
of the scattering cross section gives a `trivial' result. The essential contribution
is
\be
\sigma_{\rm tr}^{\rm hard} =
  \frac1s \int_{-s}^{t^\star} \d t \, |t| \frac{\alpha^2}{t^2}
   =  
   \frac{\alpha^2}{s} \log \frac{s}{|t^\star|}  \, ,
  \label{Y-hard}
\ee
 identical to \eq{eq1} of course.

\begin{figure}[hbt]
\includegraphics[scale=\FigScale]{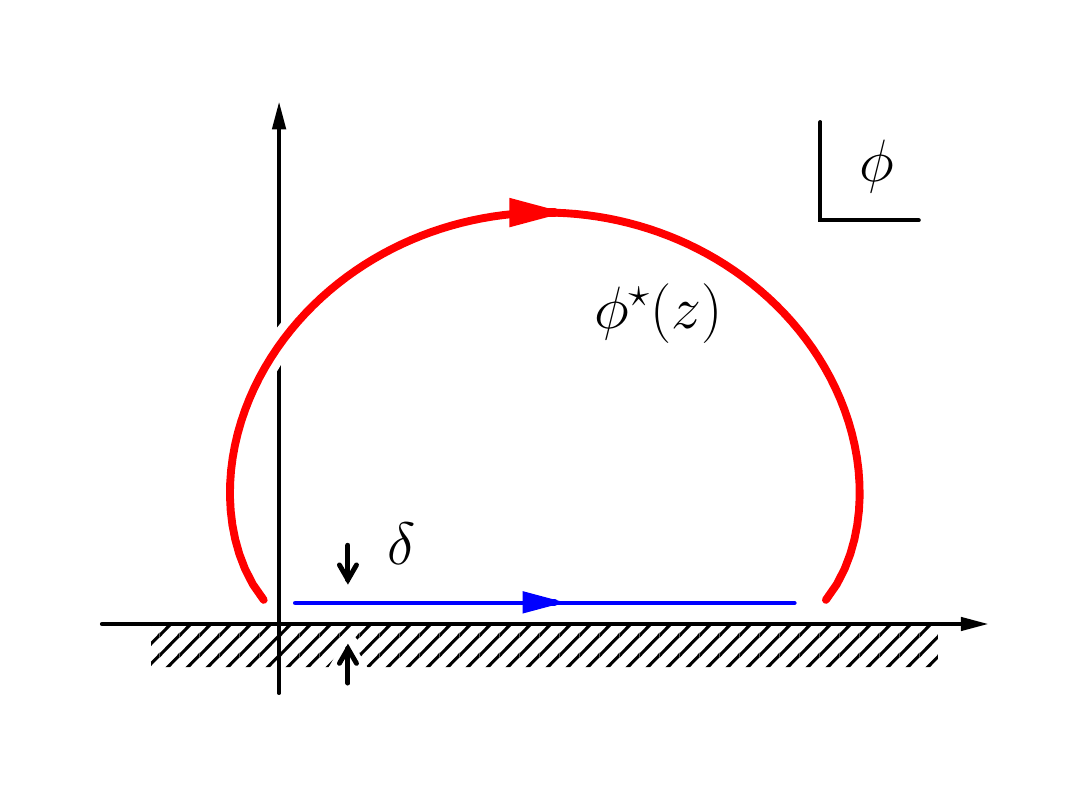}
\caption{ 
  The transport cross section takes into account screening via the function $\phi$
  (which has real and complex parts), thus tracing a curve in the complex plane with
  $z$ as the parameter.
}
  \label{fig: contour}
\end{figure}

Screening is necessary for the complementary region, where 
$t^\star < t < 0$. 
There is a further integration over the parameter
 $z \in[-1,+1]$  for time-like exchanges,
\be
  \sigma_{\rm tr}^{\rm soft}
  =
  \frac1{2s} \int_{-1}^{+1} \d z \int_{t^\star}^0 \d t \frac{(-t)\alpha^2}{| t - \wv\Pi |^2}\, ,
  \label{screen}
\ee
where $\wv\Pi = \wv\Pi_{L,T} (t,z)$ is the self energy. 
Since $|t^\star |\ll T^2$, the self-energy is independent of $t$ (so that HTL functions apply).
This expression, when combined with
the complementary region $t<t^\star$, produces no residual dependence on $t^\star$
(for weak coupling). In
other words, it tells us at which point the HTL dressing is to be replaced by tree
level functions.
The expression in Eq.~\eq{screen} was determined numerically, or in 
limiting cases. However it is possible to find the large-$|t^\star|$
contribution analytically in general. 
Keeping only relevant terms,
\bean
  \int \d t \frac{(-t)}{| t - \Pi |^2} &=&
  \int \frac{\d t}{\Im \Pi} \Im \frac{\Pi}{t-\Pi} 
  = \frac{1}{\Im \Pi} 
  \Im \Big[\, \Pi \log \frac{|t^\star| + \Pi}{\Pi} \, \Big] \, .
\eean
Making use of the fact that $|t^\star| \gg |\Pi(z)| \sim \alpha T^2$
for weak coupling. 
To extract the leading cut-off dependence, up to ${\cal O}(m_D^2/t^\star)$,
we re-express the logarithms by
\bean
\log \frac{|t^\star|}{\Pi} &=&
\log \frac{|t^\star|}{m_D^2} + \log \frac{m_D^2}{\Pi} \, ,
\eean
where $m_D^2 = 4\pi \alpha T^2$ is the LO Debye mass (for $\nf=0$).
The familiar log-structure emerges
when $| t^\star | \rightarrow \infty$, plus a constant next to
the logarithm,
\be
  \sigma_{\rm tr}^{\rm soft} =
  \frac{\alpha^2}{s} \Big( \log \frac{|t^\star|}{m_D^2}
- 
\int_{0}^{1} \d z 
 \frac{1}{\Im \phi} \Im\! \big[ \, \phi \log \phi \, \big]  \Big) \, ,
 \label{soft}
\ee
in terms of the function $\phi(z) = \Pi/m_D^2$, which is order of unity.
We have used $\phi(z)^* = \phi(-z)$, to replace the original $z$-integral
by one from $z=0$ to $1$.
A contour in the complex $\phi$-plane, parametrised by $z$,
is traced out with endpoints at $\phi(0)$ and $\phi(1)$, see Fig.~\ref{fig: contour}.

Formally, $\sigma_{\rm tr}^{\rm hard} + \sigma_{\rm tr}^{\rm soft}$ to 
NLL accuracy is then independent of $t^\star$, which cancels.
I.e. $\sigma_{\rm tr}^{\rm NLL} = \frac1s \alpha^2 [ \log (s/m_D^2) - \widetilde c\, ]$,
with $\widetilde c$ defined by the integral in \eq{soft}.
Relative errors in \eq{soft} are supressed by $m_D^2/|t^\star| \sim \alpha$ and 
amounts to a residual dependence on $t^\star$.

Equation \eq{soft} is directly applicable to the calculation for the
transport cross section. In the quenched case, one needs the
screened matrix element from $gg \rightarrow gg$ interactions. 
Take only the $t$- and $u$-channel from \eq{glue-glue}
with \eq{IR replacement}, one finds
\bea
  \frac{\d \sigma}{\d t} \Big|_{t>t^\star} &=&
  \frac{\alpha^2}{3}
  \Big[\ 
    \frac{1}{| t - \wv\Pi_L(z) |^2} + \frac{1/2}{| t - \wv\Pi_T (z) |^2}
  \ \Big] \, ,
  \label{Y-soft}
\eea
for $|t| \ll T^2$, and omitting all unnecessary prefactors.
There are two terms of the type discussed in \eq{screen}, where screening is
accounted for by HTL gluon self energies, viz.
\bean
  \phi_L (z) &=&
  (1-z^2) \Big(\, 1 - \tfrac12 z \log \frac{ z+1 }{ z-1 } \, \Big) \, ,\\
  \phi_T (z) &=&
   \Big(\,  \tfrac12 z^2 + \tfrac14 z(1-z^2) \log  \frac{z+1}{z-1}  \, \Big) \, .
\eean
Since $|z|<1$, the logarithms are complex, with values taken on the principal branch.

One has the freedom to deform
this contour to a simple one (just above the real axis), say $\phi = \gamma + i \delta$
and examine the limit $\delta \to 0$.
This dramatically simplifies the integrand appearing in Eq.~\eq{soft}; 
\be
  \Im \Big[\, \frac{ \phi \log \phi }{\delta }\, \Big]
  = \log \Big( \sqrt{\gamma^2+\delta^2}\, \Big)  
  + \frac{\gamma}{\delta} \arg (\gamma+i\delta) 
  \stackrel{\delta \to 0}{\simeq} 
  \log |\gamma| + 1 \, .
\ee
In the small-$|t|$ limit of \eq{Y-soft}, we can evaluate the transport cross section as follows.
Deform the contour, see Fig.~\ref{fig: contour}, to find that (for example) the
longitudinal propagator gives a constant in \eq{soft},
\be
\int_0^1 \d |\phi_L| \big(\, \log |\phi_L| + 1 \, \big) 
= |\phi_L(1)| \log |\phi_L(1)| 
- |\phi_L(0)| \log |\phi_L(0)| 
= 0 \, .
\ee
Only the transverse screening is non-zero, having $\phi_T(1) = \frac12$, and 
ends up giving $\widetilde c = -\frac13 \log 2$ next to the LL in \eq{soft}.

Combining the contribution from hard and soft regions given by \eq{Y-hard} and
\eq{Y-soft} produces
\bean
\sigma_{\rm tr}
  & = &
  \frac{\alpha^2}{s}
  \Big( \, 
   \log \frac{ s }{m_D^2 } - \tfrac13 \log 2 \, \Big) \, .
\eean
This demonstrates the Braaten-Yuan (BY) method \cite{Braaten1991d}, 
and confirms an earlier
numerical result of Heiselberg in Eq. (B7) \cite{Heiselberg1994b}. 
If we had used this result to match $\mu^2 = \kappa \cdot m_D^2$ in the 
simple model \eq{m1}, we would have found $\kappa = 2^{1/3}/e \approx 0.46$
[this differs from the value $\kappa \approx 0.62$ obtained via \eq{eta NLL}, due to the 
omission of inelastic proccess, the gluon 4-vertex and $s$-channel diagrams in ${\cal M}$].
Incidently, this technique is applicable for calculating the photon rate and
proves a result in \cite{Kapusta1991} previously known only numerically.
Formally, $\sigma_{\rm tr}$ is then independent of $t^\star$ in the BY scheme resting on the
assumption that $m_D^2 \ll -t \ll s \sim T^2$.
Residual dependence on $t^\star$ in \eq{Y-soft} for
subleading  errors that are $\sim \Pi/|t^\star|$ and in \eq{Y-hard} they are $\sim |t^\star|/s$.

Let us end with a comment on the finite radius of convergence for $\sigma_{\rm tr}$.
In Sec.~\ref{sec 2A} we found that it was simply the `mass' $\mu^2$ -- here it
will instead be given by the minimum of $|\Pi(z)|$.
That is zero for HTL functions, but in general may be related to the QCD `magnetic' mass.
This speculation is well beyond our present scope.

%% file: appendix-3.tex
\renewcommand{\arraystretch}{1.7}
\section{Two-body phase space\label{A2}}

The collisional operator \eq{C[f]} expresses the rate of binary encounters, integrated
over partner momenta $\bm p_2$, $\bm p_3$ and $\bm p_4$.
For a given function $g \big( \{ \bm p_i \}\big)$ which depends on the participant
momenta, we evaluate the phase space integral,
$\int \d \Gamma \cdot g$.
One of the integrals may be completed using energy-momentum conservation; 
we choose the $p_4$-integral, 
\bea
\int_4 \frac{g}{2E_4} &=& 2\pi 
\delta ( \underline{K}_4^2 ) \theta (\underline{E}_4) \ \underline{g} \, .
\label{B2}
\eea
Here the underline indicates a dependence on the fixed 4-momentum 
$\underline{K}_4 = P_1 + P_2 - P_3$. For instance, $\underline{f}_4$ depends on the
energy $\underline{E}_4 = E_1 + E_2 - E_3$. It will turn out that the on-shell
constraint $\underline{P}_4^2 = 0$, expressed by the $\delta$-function in \eq{B2}, 
implies already $\underline{E}_4 \geq 0$, making
the factor of $\theta (\underline{E}_4)$ redundant in \eq{B2}. 
The remaining fivefold integral
is further reduced as follows \cite{Peigne2008}. We align the $z$-axis with $\bm p_1$
and orient the $zy$-plane to contain $\bm p_3$ viz.
\bea
\bm p_1 &=& E_1 (0,0,1) \, , \nonumber \\
\bm p_2 &=& E_2 (\sin \phi \sin \theta_2, \cos \phi \sin \theta_2, \cos \theta_2 ) \, ,
\nonumber \\
\bm p_3 &=& E_3 (0, \sin \theta_3, \cos \theta_3 ) \, .
\label{p123}
\eea
The argument of the $\delta$-function in \eq{B2} depends on $\phi$ through
\bean
\underline{P}_4^2 &=& 2(P_1 P_2 - P_1 P_3 - P_2 P_3 ) \\
&=& s + t - 2 P_2 P_3 \, .
\eean
Thus, using \eq{p123} to simplify the outstanding 4-product, we have
$\underline{P}_4^2 = A + B \cos \phi$ where 
\bea
A &=& s + t -2 E_2 E_3 (1-\cos \theta_2 \cos \theta_3 ) \, ,\nonumber \\
B &=& 2 E_2 E_3 \sin \theta_1 \sin \theta_3 \, . \label{AB}
\eea
Emphasising the azimuthal dependence in $g = g (\phi)$, the $\phi$-integral is elementary
\bean
\int_0^{2\pi} \d \phi\ 
\delta \l( \underline{P}_4^2 \r)
\ g(\phi)
&=&
E_1 \frac{ \theta (h) }{\sqrt{h}} \sum_{\pm} g(\phi_\pm) \,
,
\eean
where $\phi_\pm = \pi \pm \arccos(A/B)$ and $h = E_1^2 ( B^2 - A^2 )$ 
($E_1$ is factored out for convenience).
Next, we reformulate the remaining integration over 
 $\cos \theta_{2,3}$ in terms $s$ and $t$, whose values are specified by $\bm p_1,
\bm p_2$ and $\bm p_3$:
\bea
s &=&  2 E_1 E_2 (1 - \cos \theta_2 ) \, , \nonumber \\
t &=& -2 E_1 E_3 (1 - \cos \theta_3 ) \, , \label{st}
\eea
with Jacobian $4E_2E_4 \cdot E_1^2$. Using this, and \eq{B2}, we arrive at
\bea
\int \d \Gamma \cdot g(\cdots) &=&
\frac{1}{16 (2\pi)^4 E_1} \int \d s \d t \ \int \d E_2 \d E_3 
\frac{\theta (h)}{\sqrt{h}} \sum_\pm \underline{g} (\phi_\pm) \, .
\label{B3}
\eea
The function $h$ is quadratic in $E_3$, written as $h(E_3) = aE_3^2 + bE_3+c$ the
coefficients are found from \eq{AB} and \eq{B3} to be
\bean
a &=& -s^2 \, ,\\
b &=& -2s(uE_1+tE_2) \, ,\\
c &=& -(uE_1 - t E_2)^2 - stu \, .
\eean
Since $a<0$,  $\theta \l( h \r)$ constrains the $E_3$-integration to the
interval $[E_3^-, E_3^+]$, where
\bean
  E_3^\pm &=& \frac{-b \pm \sqrt{D}}{2a} \, .
\eean
Positivity of the discriminant $D =4 s^2 t u (4 E_1 E_2 - s)$ summarises the 2-body
phase space: 
$0 \leq s \leq s_{\rm max} = 4 E_1 E_2$
and $-s \leq t \leq 0$. 

With the kinematic bounds fully specified, the factor $\theta (h)$ can indeed (as
anticipated)  be dropped in
\eq{B3} to give
\bea
\int \d \Gamma \cdot g &=&
\frac{1}{(2\pi)^3 E_1} 
\int \d E_2
\int_0^{s_{\rm max}} \d s 
\int_{-s}^0 \d t \ 
 \frac{1}{16\pi}{\cal I} (s,t ,E_1, E_2) \, ,
\label{stuway}
\eea
in terms of a function that parametrises the final $E_3$-integral:
\bea
 {\cal I} [g] = 
\int_{E_3^-}^{E_3^+} \frac{ \d E_3 }{\sqrt{ h(E_3) }}
\ \frac12\sum_\pm \underline{g} ( E_3, \phi_\pm ) \, .
\label{K}
\eea
Up to this point, there have been no simplifying approximations. 
After using energy and momentum conservation as well as symmetry in
one of the angles, we have reduced the original expression \eq{B2}
to a four dimensional integration.

\subsection*{Small-$\bm t$ approximation}

Of interest to Sec.~\ref{sec:sigTR}, is the behaviour of \eq{stuway} assuming dominance of
small angle scatterings.
We now discuss in some more detail how to complete the $E_3$-integral of \eq{stuway},
keeping terms of relevant powers in $t$.
For small $\omega = E_1 - E_3$, it is reasonable to replace in $g(E_3)$ the energy
$E_3 \to E_1$. 
However, for $-t \ll T^2$ it remains possible that both $|\omega|$ and $q$
are individually large.

The integrand \eq{K} has nonzero support for $h(E_3) > 0$, which as $|t| \to 0$ becomes
very narrow; $(E_3^+-E_3^-) = \frac2s \sqrt{tu(s_{\rm max}-s)}$ with the central
value
\bean
E_3^\star &=& E_1 - \frac{t}{s} \big( E_2-E_1 \big) \, .
\eean
Let us suppose that the integrand $g$ is largest for small-$t$ (or $u$), and
accordingly expand $g(E_3,\cdots)$ about $E_3^\star$ (\ie\ if $-t\ll s$, then $E_3 \simeq E_1$).
We thus represent $g$ in powers of $(E_3 - E_3^\star)$, to be integrated in \eq{K},
leading us to 
\bea
{\cal I} [g]  = 
\sum_{k=0}^\infty {\cal I}_{(k)} 
\frac{\pd^{k} g}{\pd E_3^{k}} \bigg|_{E_3=E_3^\star} \, ,
\label{fullK}
\eea
where ${\cal I}_{(k)} = {\cal I}[(E_3-E_3^\star)^k]/(k!)$.
Only the first few terms in \eq{fullK} turn out to be relevant, namely
\be
\begin{tabular}{c||c|c|c}
  $k$  &  0  & 1 &   2  \\
  \hline
  $ {\cal I}_{(k)}\  $
  & $\ \D \frac{\pi}{s} \ $
  & $\ \D 0 \ $
  & $\ \D \frac{\pi}{16} \frac{D}{s^5} \ $ 
\end{tabular}  \, ,
\label{K_(m)}
\ee
for $k\geq 3$ we find ${\cal I}_{(k)} = {\cal O}(D^{k/2})$,
which is subleading for small-$t$ or $u$.
If we let $g^\prime = \pd g / \pd E_3$, 
then the leading terms for a small-$t$ approximation may be written
\bea
{\cal I}[g] &\simeq& 
\frac{\pi}{s} g(E_3^\star)
  + \frac \pi{4s^3} t u \big( 4E_1E_2 - s \big) g^{\prime\prime} (E_3^\star) 
\label{E3int}
\eea
It will turn out (see below), basically because of the choice of tensor basis \eq{chiij}, that 
$g(E_3^\star) \propto tu$ and therefore
 both terms above are ${\cal O}(tu/s^2)$\,.
The pertinent transport weight, $|t|$ in \eq{E3int}, emerges from the peaked kinematics for $E_3\sim E_1$
as $t\to 0$ (or $E_3 \sim E_2$ as $u\to 0$).

\bigskip

For the linearised collisional operator \eq{symC}, the QCD matrix element 
 is largest when $t$ or $u$ is small compared with $s$.
Here the integrand is
$$g(\cdots) = |{\cal M}|^2 f_1 f_2 \bar f_3 \bar f_4  \cdot \big( \Delta^{ij}[\chi] \big)^2 \, ,$$
which depends on the unknown function $\chi$.
Because $E^\star_3 \simeq E_1$, up to subleading corrections for $-t\ll s$, we can replace $f_3 \to f_1$,
which gives
\bea
\int \d \Gamma \cdot g &=& 
\frac{f_1 \bar f_1}{(2\pi)^3 E_1} 
\int \d E_2\, f_2 \bar f_2 
\int \d s \, s^2 
\int \d t \frac{\d \sigma}{\d t} \,
{\cal I}\bm[\, \big( \Delta^{ij}[\chi] \big)^2 \,\bm] \, ,
\label{g's}
\eea
 assuming that the cross section $\d \sigma/ \d t = |{\cal M}|^2/(16\pi s^2)$
is only a function of $s$ and $t$.
$\chi$ represents the input to the functional ${\cal Q}[\chi]$, see Eq.~\eq{Q}, whose maximal value
gives $\eta$. Hence
$$
{\cal I} = \frac{\pi |t|}{4s^2}  
\big[ \, 8 s {\cal A} + \big( 4E_1E_2 - s \big) {\cal B} \, \big] \, ,
$$
where ${\cal A}$ and ${\cal B}$ were given in \eq{Pdef}.